\documentclass[5p,times]{elsarticle}
\usepackage{hyperref}
\usepackage{graphicx}
\usepackage{tikz}
\usepackage{tikz-qtree}
\usetikzlibrary{shapes}
\usepackage{longtable}
\newcommand{\ts}{\textsuperscript}

\journal{Journal of Information and Software Technology}

\begin{document}

\begin{frontmatter}

\title{Software Sustainability: A Systematic Literature Review and Comprehensive Analysis}

\author[mymainaddress]{Asif Imran}


\cortext[mycorrespondingauthor]{Asif Imran}
\ead{asifimra@buffalo.edu}

\author[mymainaddress]{Tevfik Kosar}
\ead{tkosar@buffalo.edu}

\address[mymainaddress]{Department of Computer Science \& Engineering \\ University at Buffalo (SUNY), Buffalo, NY, 14260, USA}

\begin{abstract}
\textbf{Context:} Software Engineering is a constantly evolving subject area that faces new challenges everyday as it tries to automate newer business processes. One of the key challenges to the success of a software solution is attaining sustainability. The inability of numerous software to sustain for the desired time-length is caused by limited consideration given towards sustainability during the stages of software development.\\ 
\textbf{Objective:} This review aims to present a detailed and inclusive study covering both the technical and non-technical challenges and approaches of software sustainability.\\
\textbf{Method:} A systematic and comprehensive literature review was conducted based on 107 relevant studies that were selected using the Evidence Based Software Engineering (EBSE) technique.\\
\textbf{Results:} The study showed that sustainability can be achieved by conducting specific activities at the technical and non-technical levels. The technical level consists of software design, coding, and user experience attributes. The non-technical level consists of documentation, sustainability manifestos, training of software engineers, funding software projects, and leadership skills of project managers to achieve sustainability. This paper groups the existing research efforts based on the above aspects. Next, how those aspects affect open and closed source software is tabulated. Based on the findings of this review, it is seen that both technical and non-technical sustainability aspects are equally important, taking one into contention and ignoring the other will threaten the sustenance of software products.\\
\textbf{Conclusion:} Despite the noteworthy advantages of making a software sustainable, the research community has presented only a limited number of approaches that contribute to sustainability. To the best of our knowledge, those representations require further research. In this regard, an organized, structured and detailed study on existing technical and non-technical sustainability approaches is provided here which will serve as a one-stop-service for researchers and software engineers who are willing to learn about those.
\end{abstract}

\begin{keyword}
\texttt software sustainability\sep software engineering \sep sustainable design \sep software longevity
\end{keyword}

\end{frontmatter}


\section{Introduction}
\label{intro}
Achieving software sustainability is an important area of software engineering research today. The goal of software sustainability engineering is to ensure that software continues to achieve its goals despite updates, modifications and evolution. We follow the definition of sustainability provided by the Software Sustainability Institute which states \textit{"software you use today will be available - and continue to be improved and supported in the future"} . Other definitions of software sustainability consider the age of software and social aspects. In this paper, we consider the definition mentioned above.

Software sustainability can help us achieve a number of useful goals. Some notable goals of software sustainability are mentioned below.

\begin{itemize}
    \item \textit{\textbf{Operational efficiency}}: Sustainability of software used both in industries and by individuals should be a natural part of overall performance management practice. If the software possess the capability to sustain for a long time, there is no need to train researchers on a new type of software \cite{penzenstadler2011teach}. Also, the researchers will become more efficient if they use the same software for a long time, thereby increasing their operational efficiency \cite{penzenstadler2011teach}. Also, an individual using a software for a significant time is likely to stick to that software rather than moving to a new one.
    
    \item \textit{\textbf{Desirable reputation of software product}}: To remain competitive, companies need to make innovation one of their top priorities \cite{dagli2008system}. As an example, if the software developed by a company is both technically and non-technically sustainable, they can state that their software are long lasting and ensure high quality output \cite{durdik2012sustainability}. Hence, consumers will find the software reliable and have more confidence in using it. This, in turn will provide the company with the capacity of building a desirable reputation.
    
    \item \textit{\textbf{Reduced cost}}: If the software which is used by an industry or individual users for day-to-day activities is technologically sustainable, then that industry or individual does not need to invest on a new software in the near future, thereby their capital expenditure is reduced unless a new software is procured which offers increased benefits and better fits the business needs \cite{Stewart:2015:SSC:2753524.2753533}. On the other hand, if the software is not sustainable and it needs to be replaced within a short time, the users (both industry and personal) need to spend on procuring the new software, installing it on the computers, arranging for training on the use of the new software, etc. Hence, both the capital and current expenditures will rise due to the lack of sustainability of software \cite{seacord2003measuring}. From a business perspective, investing on a software which is sustainable, will guarantee cost reduction and profit increase in the long run \cite{seacord2003measuring, Chitchyan:2016:SDR:2889160.2889217}.
    
    \item \textit{\textbf{Accelerated progress of scientific software}}: The influence of digital technology in modern research is manifold, where data and publications are being produced, shared, analyzed and stored using various types of scientific software. Although research software plays an important role in the field of science, engineering and other areas, in most cases they are not developed in a sustainable way. The researchers who develop them, maybe well-versed in their own discipline, however, they may not have required knowledge on the best practices of software maintainability and sustainability which are needed for reproducibility of simulation results. As stated by \textit{US Research Software Sustainability Institute (URSSI)}, there is a need for strategic plan that will conduct the necessary activities like training, prototyping, implementations with a goal to create improved, more sustainable software \cite{urssi}. This software in turn will accelerate the progress of science.
\end{itemize}

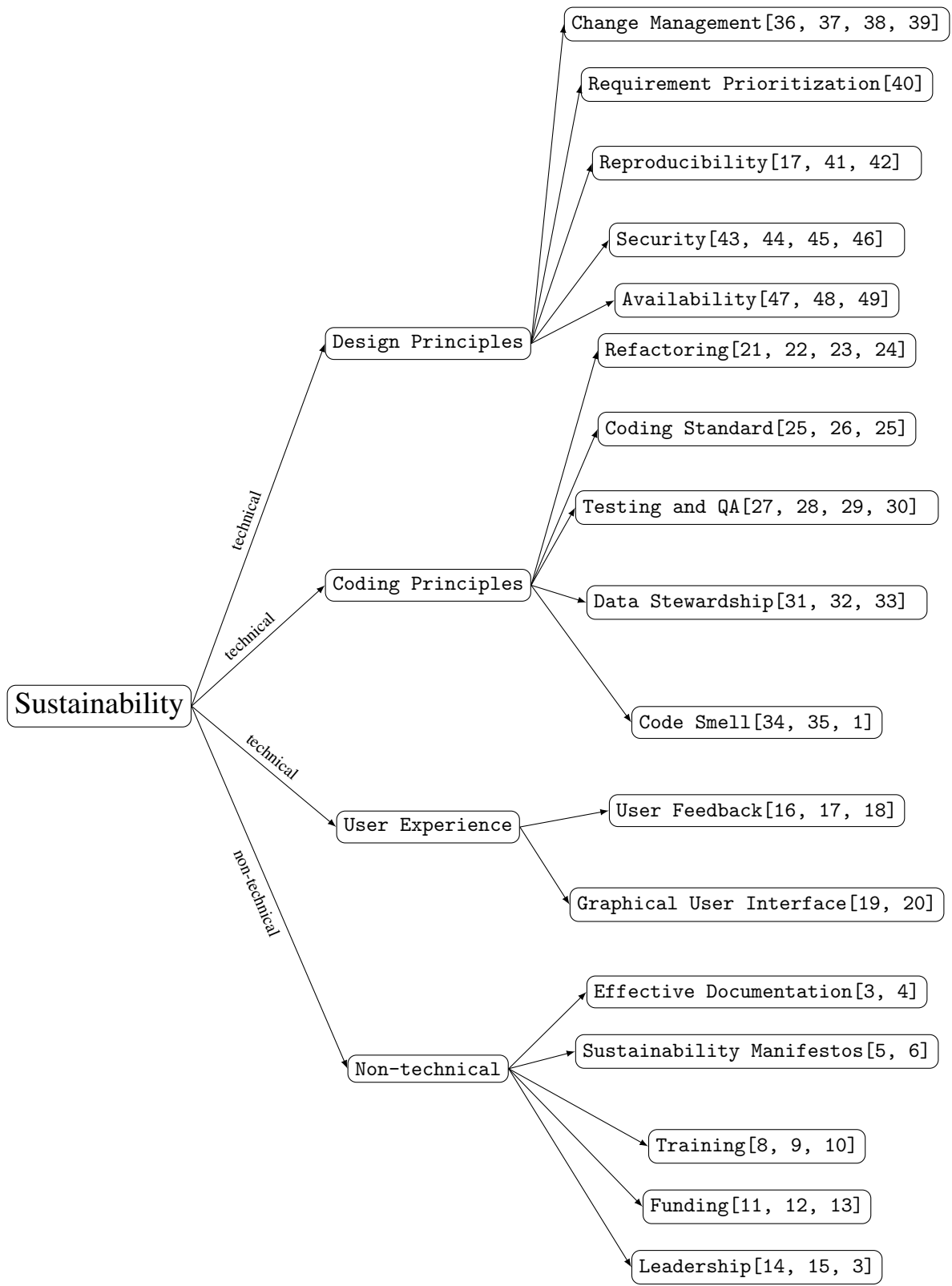
\begin{figure*}
\centering
\tikzset{
  treenode/.style = {shape=rectangle, rounded corners,
                     draw, align=center,
                     top color=white, bottom color=white!20},
  root/.style     = {treenode, font=\Large, bottom color=white!30},
  env/.style      = {treenode, font=\ttfamily\normalsize},
  dummy/.style    = {circle,draw}
  edge from parent path={
(\tikzparentnode) |-   
($(\tikzparentnode)!0.5!(\tikzchildnode)$) -| 
(\tikzchildnode)}] 
}
\begin{tikzpicture}
  [
    grow                    = right,
    sibling distance        = 4cm,
    level distance          = 5.4cm,
    edge from parent/.style = {draw, -latex},
    every node/.style       = {font=\footnotesize},
    sloped
  ]
  \node [root] {Sustainability}
    child { node [env] {Non-technical}
      child { node [env][above=8cm] {Sustainability Manifestos\cite{Becker:2015:SDS:2819009.2819082,Ojameruaye:2016:SDP:2889160.2889218}} }
      child { node [env][above=5cm]{Effective Documentation\cite{katerbow2018recommendations,hutton2016most}} }
      child { node [env][below=1cm] {Training\cite{brown2002reusing,mozilla.grants,druskat2018mapping}} }
      child { node [env][below=6cm] {Funding\cite{druskat2016lightning,cohen2018building,8257959}}}
      child { node [env][below=11cm] {Leadership\cite{Stewart:2015:SSC:2753524.2753533,plotnick2008leadership,katerbow2018recommendations}}}
      edge from parent node [below] {non-technical} }
    child { node [env] {User Experience}
      child { node [env][above=02cm] {User Feedback\cite{da2000user,kitzes2017practice,gilb1988principles}} }
      child { node [env][below=03cm] {Graphical User Interface\cite{shneiderman2010designing, tidwell2010designing}}}
      edge from parent node [above] {technical} }
    child { node [env] {Coding Principles}
      child { node [env][above=11.6cm] {Refactoring\cite{8443579,hdf.2018,o'neal_weide_dubey_2018,amrex}}}
      child { node [env][above=6.3cm]{Coding Standard\cite{petcu2011towards,zervas2017rise,petcu2011towards}}}
      child { node [env][above=1cm] {Testing and QA\cite{robillard2016sustainable,tsantalis2006design,petre2013software,jetuml} }}
      child { node [env][below=4cm] {Data Stewardship\cite{abadi2009data,Chang:2008:BDS:1365815.1365816,6982633} }}
      child { node [env][below=10cm] {Code Smell\cite{Palomba:2015:TAC:2819009.2819162,azeem2019machine,Koziolek:2011:SES:2000259.2000263}}}
      edge from parent node [above] {technical} }
    child { node [env] {Design Principles}
      child { node [env][above=13cm] {Change Management\cite{staahl2018dynamic,atomic,da2014empirical,anvik2006should}}}
      child { node [env][above=8cm] {Requirement Prioritization\cite{katz2015report}}}
      child { node [env][above=2.7cm] {Reproducibility\cite{kitzes2017practice,jimenez2017popper,staubitz2016codeocean} }}
      child { node [env][below=2cm] {Security\cite{penzenstadler2014safety,Lago:2015:FSP:2830674.2714560,norris2004mission,Guessi:2015:SLR:2695664.2695795} }}
      child { node [env][below=7cm] {Availability\cite{Marwah:2010:QSI:1773394.1773405,tokuno2003relationship,lago2010service}}}
      edge from parent node [above] {technical} };
\end{tikzpicture}
\caption{Different aspects of software sustainability} \label{fig:M1}
\end{figure*}

There are different kinds of activities involved in software sustainability engineering. Depending on their complexity and applications, conceptually, sustainability can be divided into two broad levels: technical and non-technical as shown in Figure \ref{fig:M1}. 
Technical category includes the software \textit{design principles, coding principles, and user experience} frameworks which help to achieve sustainability \cite{8500171}. Software \textit{design principles} include the standards deployed to organize the structural components of software engineering \cite{1223642}. Those principles are applied from the beginning of software development life cycle till the end. \textit{Coding principles} are the mechanisms of structuring new code or restructuring older code with an aim to improve longevity \cite{7337457}. \textit{User experience} is the set of design practices which tailor the user experience to match the needs and expectations of the users with regard to their use cases, hence improving their satisfaction which causes them to persist with the software for a longer time \cite{da2000user}.
\\

Non-technical category of software sustainability engineering involves the non-functional aspects, which are shown in Figure \ref{fig:M1}. Those include \textit{effective documentation, sustainability manifestos, training of software engineers} on how to build sustainable software, \textit{funding} to integrate sustainability characteristics, and \textit{leadership} capabilities of the software project manager to achieve sustainability. \textit{Effective documentation} ensures that the software documents like requirements document, design document, test document, and maintenance document create clear and precise communication among the development team and other stakeholders \cite{druskat2016lightning}. \textit{Sustainability manifestos} identify the policies which should be followed by a software to ensure sustenance \cite{Becker:2015:SDS:2819009.2819082}. \textit{Training} of software engineers is the process of learning to understand the newer and more sustainable software development frameworks and methodologies \cite{penzenstadler2011teach}. \textit{Funding} is the sum of money which is provided to software teams to build more sustainable software \cite{Stewart:2015:SSC:2753524.2753533}. It can be through grants, crowd-sourcing, or start-up funding. \textit{Leadership} includes the skills through which a project manager handles change management, timeline management, cost management, employee turnover, etc \cite{Stewart:2015:SSC:2753524.2753533}. We summarize all of the factors contributing towards software sustainability in Figure~\ref{fig:M1}, along with the references to the identified literature which we discuss later in this paper.

A previous survey paper by Penzenstadler et al.~\cite{penzenstadler2012sustainability} covered several low-level components for understanding software sustainability, focusing mostly on the non-technical attributes. However, there is a need to identify both the technical and non-technical attributes of sustainability together to provide a holistic viewpoint for the software engineers and researchers. Their extended review on sustainability \cite{penzenstadler2014safety} focused on the sustainable design of software. They stated that secured software design and testing are important to develop sustainable software. They limited their work to specific programming components such as commenting code and following coding standards. However, the authors did not analyze the effect of important technical aspects like requirement prioritization, code smell detection, change management, etc which equally play a role in ensuring sustainability. Calero et al. \cite{calero2013systematic} provided a review on software sustainability which is primarily based on environmental friendliness of software. However, the non-technical aspects of environmental sustainability covered in \cite{calero2013systematic} are not sufficient to represent the holistic activities needed for software sustainability. Other non-technical aspects like documentation, sustainability manifestos, funding, and leadership skills of the project manager were not considered.

The objective of this paper is to provide a systematic and comprehensive overview of the state-of-the-art software sustainability methodologies currently in place. We discuss various types of approaches designed both for technical and non-technical activities which aim to make software sustainable. We compare and contrast how these approaches apply to open source and closed source software. Specifically, open source software is software with source code accessible to anyone for the purpose of inspection, modification, and enhancement . Open source software can leverage the additional capacities and intelligence of software experts who are outside of a particular organization . On the other hand, when the source code of a software is accessible to only one person, team, or organization which maintains exclusive control of modifying the code, then it is called closed source software . Only the original authors of closed source software are allowed to legally copy, inspect, and modify the source code. In order to use the software, users must agree to make a payment for it and sign a license . The license prevents users from making any changes to the software which the software's authors have not explicitly permitted . Our findings show that open source software provides a greater window of opportunity to apply latest innovative techniques of sustainability at the design, coding, and user experience levels mainly because of its open model of contribution. However, most of the sustainability techniques are rarely applied to open source software due to lack of funds required for those. Currently, organizations like \textit{US Research Software Sustainability Institute (URSSI)}  and \textit{The Software Sustainability Institute} are investigating to address those issues. Hence extensive research is required to solve this issue. On the other hand, closed source software apply well-established but old techniques which are discussed later in the paper.

Based on our findings, we argue that under present circumstances there is still room for improvement in the field of sustainable software development. The open issues emerging from this study will provide input to researchers who are willing to develop improved techniques for software sustainability. We conclude that to achieve sustenance, both the technical and non-technical factors need to be considered simultaneously. Considering one and ignoring the other will not provide long term sustenance of the software. As a result, we review and present related research on both technical and non-technical aspects of sustainability in this paper.

The rest of the paper is organized as follows. Section \ref{methodology} provides a systematic and methodological way of identifying, assessing, and analyzing published research papers within the scope of this study. Section \ref{design} focuses on the useful literature regarding design aspects of software sustainability. Section \ref{coding} highlights the related research which has been conducted on coding practices for sustainable development. The related research showing user experience aspects which need to be considered to achieve long term sustainability of the software are presented in Section \ref{ux}. Current research focusing on non-technical aspects which are necessary for sustainability are highlighted in Section \ref{nt}. Finally, in Section \ref{conclude}, we conclude the review, summarize our findings and highlight the scope for future research in this area.

\section{Methodology of Study}
\label{methodology}
In this section, we identify the research aim and questions. We highlight the paper search method, paper filtering mechanism, thematic grouping of papers and evaluation of the review protocol. Additionally, we provide the sources from which the articles were obtained and tabulate the number of articles selected from specific sources.

\subsection{Research Aim and Questions}

The aim of this research is to identify what related bibliographic databases report regarding the challenges to software sustainability. At the same time, which of those challenges belong to technical or non-technical aspects. It is highly important to identify the sustainability challenges in order to address those. Here we aim to identify those challenges. Also, causal chaining between the various sustainability aspects need to be set up. The effect of sustainability challenges on open and closed source software need to be evaluated as well. More specifically, the following research questions are addressed.  
	
\begin{figure*}
  \centering
  \includegraphics[width=17cm,height=8.6cm]{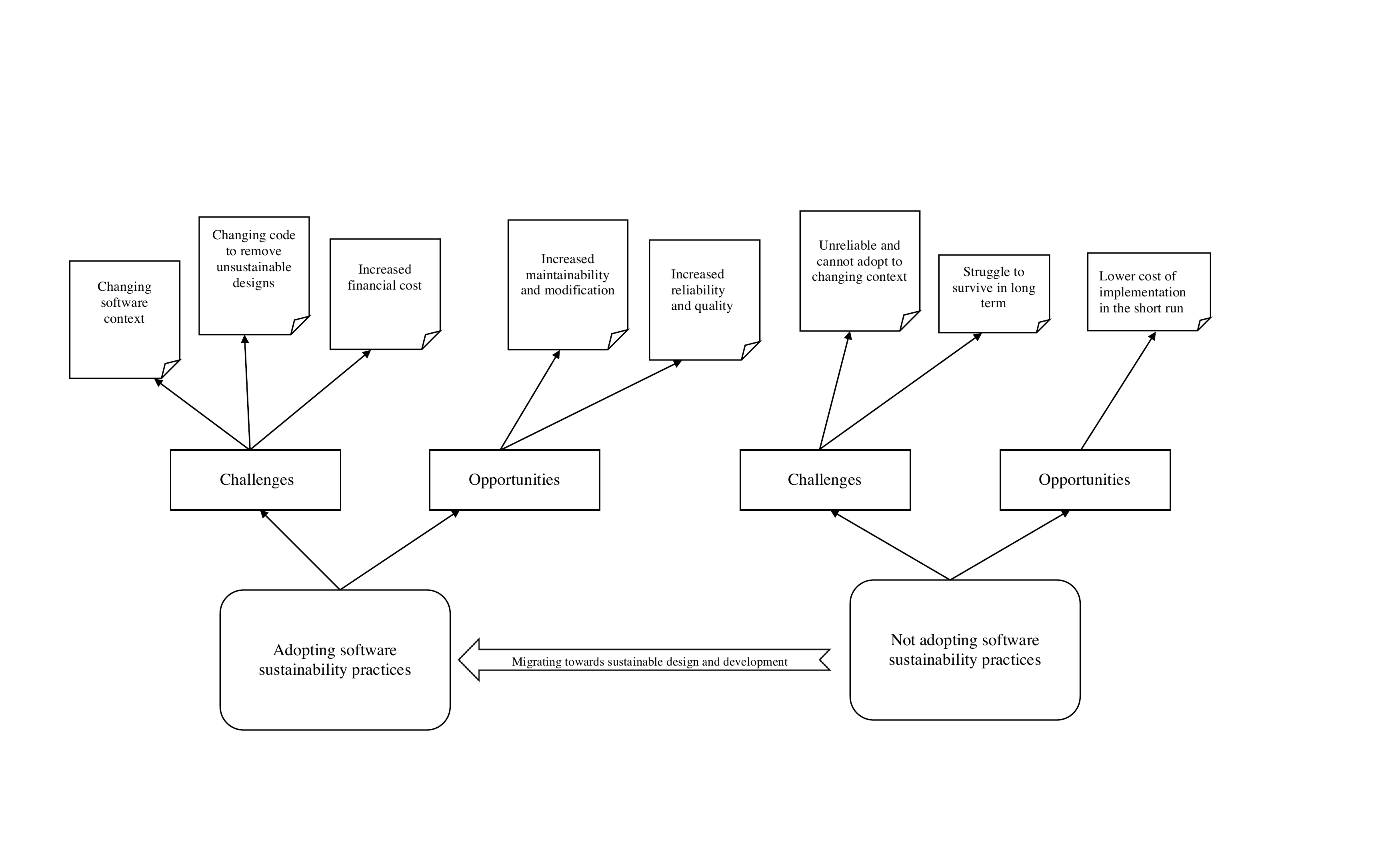}
  \caption{Challenges and opportunities of adopting software sustainability practices}
\label{fig6}
\end{figure*}

\begin{itemize}

\item RQ1: What technical and non technical challenges to software sustainability have been identified in leading research databases?

\item RQ2: What causal relationships exist between the sustainability challenges and software engineering practices?

\item RQ3: How the identified challenges vary between open and closed source software?

\end{itemize}
\begin{table}
\centering
\begin{tabular}{ |p{4cm}||p{2cm}|p{2cm}|  }
 \hline
 \multicolumn{3}{|c|}{Total number of articles before applying exclusion criteria} \\
 \hline
 Bibliographic Database     & Pre 2010 & Post 2010\\
 \hline
 IEEE Explore  & 44    &60\\\hline
 ACM Digital Library&   15  & 64  \\\hline
 ScienceDirect &10 & 30\\\hline
 ISI Web of Science  &5 & 18\\\hline
 Springer&   20  & 4\\\hline
 Google Scholar& 25  & 37\\\hline
 URSSI& 0  & 10\\\hline
 SSI& 0  & 16\\\hline
 Scopus& 10  & 23\\
 \hline
\end{tabular}
\caption{Search results for each database, divided in two groups of pre 2010 and post 2010}
\label{article}
\end{table}
These research questions questions are identified based on the PICOC (Population, Intervention, Comparison, Outcome, Context) criteria identified by Petticrew et al.which was also adopted by Kitchenham et al \cite{kitchenham2013systematic}.  The target \textit{population} of this study covers researchers and software engineers concerned with sustainability of software. The \textit{intervention} includes the technical and non-technical aspects and challenges of software sustainability. \textit{Comparison} can be done with the attributes and vulnerabilities of a software which is not sustainable such as suffering from community smells, design and code smells . Figure \ref{fig6} gives a pictorial analysis of the comparison. \textit{Outcomes} include calculation of improvement in maintainability, reliability, wider user-acceptance, and greater longevity of sustainable software as provided in existing articles. Finally, \textit{context} covers industry which focus on developing real-life sustainable software, as well as researchers who focus on finding techniques for the purpose of mitigating the sustainability challenges. The above factors were considered during determination of the research questions to ensure greater impact.

We aim to answer the above question through a systematic literature review of technical and non-technical sustainability challenges as identified in related research articles. This study is limited to the leading bibliographic databases as identified in Table \ref{article}. It includes research articles in terms of papers published in journals, conferences, symposiums, workshops, etc. It also includes technical reports and book chapters. However, blog posts, video and audio tutorials, slide shows have not been included in this study. Despite this exclusion, the articles considered here make a good contribution since those belong to the major bibliographic databases and are thus considered reliable. Also, we aimed to limit the scope of papers which abide the definition of sustainability provided by Software Sustainability Institute and identified earlier in this paper. In addition to the major databases, we included articles published in the workshops of two notable sources namely \textit{Software Sustainability Institute}and \textit{US Research Software Sustainability Institute} \cite{urssi}. The reason for this consideration is that the articles published at those workshops are highly related to the scope of this study. 

The review protocol for this study was determined based on the guidelines of Kitchenham et al.\cite{kitchenham2013systematic}. The strategy included search terms which were determined from the research questions. Next, the search domain was determined which included popular and major bibliographic databases. The selected articles were reviewed by the authors and disagreements were resolved using Cohen's Kappa as described later. The data extraction strategy included the use of extraction forms provided in Kitchenham et al. \cite{kitchenham2013systematic} to collect all relevant information in selected articles. Extracted data were presented in this study using a thematic pattern of technical and non-technical aspects of sustainability. Multiple articles from the same authors were included in the study if found relevant, however, articles whose data and results matched a previously selected paper were discarded. The overall review protocol consisted of the following stages.

The preliminary search described in the following section resulted in many articles which included the keyword sustainability, however were not relevant to this study and neither belonged within the scope of the definition of software sustainability considered here. Hence, upon suggestions from the second author, advanced searches were conducted as specified in Kitchenham et al. \cite{kitchenham2013systematic}with various combinations of keywords obtained from research questions which was broken down into facets using PICOC criteria. Next, a check was conducted by the authors between the advanced and preliminary search lists. The common papers were considered for further evaluations. Following is a description of the search strings and deployed search strategy.

\subsection{Search Mechanism}

We have used the methodology of \textit{Evidence Based Software Engineering} (EBSE) \cite{kitchenham2013systematic} to identify the relevant papers for this topic. First, we conducted preliminary search in using string derived from research questions like \textit{"software + sustainability + engineering", "challenges + software + sustenance", "non-technical + aspects + software + sustainability", "software + sustainability + open + source"}, etc. Next, we determined the categories based on the principle of \textit{Software Development Life Cycle} (SDLC) \cite{kitchenham2013systematic}. The term \textit{"software + engineering + sustainability"} was used to ensure that the extracted data properly address the research questions. The search strings were applied to the title, abstract, keywords, and conclusions of the articles. The first search was conducted in March 2018 and the last search was conducted in early September 2019. The later search was conducted to include papers which may have been published post the initial search date. 

Once the above conditions were satisfied, we finalized our taxonomy for selection of research papers. Based on the manual search on categories, we selected candidate papers focusing on their abstract, title, keywords, and conclusion. Thus, an initial list of papers became available. Afterwards, the papers in the initial list were collated which resulted in an initial list of 391 articles. Next, we used the filtering mechanism which is discussed in the next sub-section. After the filtering process, forward \textit{snow-balling} \cite{petersen2015guidelines} technique was applied , which is an automated method to search for papers which cited the ones in our initial list. From the snowball list, the process of selecting candidate papers based on the title, abstract and keywords were again repeated. Once the papers were agreed to be included, the full versions of the papers were obtained and read during the data extraction procedure. We recorded the crucial contributions, results and limitations in those papers. This formal process reduced the chances of any prolific papers being missed in this study. Repeating the search with backward snowballing was not necessary in the context and topic of this research as most of the influential papers were obtained by forward snowballing technique. This is because both backward and forward snowballing rely on title of the paper and the list of references. Hence, after selecting the articles and applying backward snowballing, any additional studies could not be identified.

\subsection{Filtering Mechanism}
The filtering process used in this study is valuable since it filters the major articles with respect to the research questions using an unbiased search strategy from popular bibliographic databases. Filtering research articles is a multi-step process including a series of inclusions and exclusions \cite{kitchenham2013systematic}. Initially, a liberal approach was taken such that if an article cannot be excluded solely based on title, abstract, keywords, and conclusion, the full text was obtained. Conclusion was analyzed since Brereton suggested that standard of IT and software engineering abstracts may be sometimes misleading to select the article for review. 

The guide which was used to filter the research papers included the following criteria:

Factual: A paper discussing technical and non-technical software sustainability needs to do so within the scope of the definition followed here.

\begin{figure*}
  \centering
  \includegraphics[width=16cm,height=8cm]{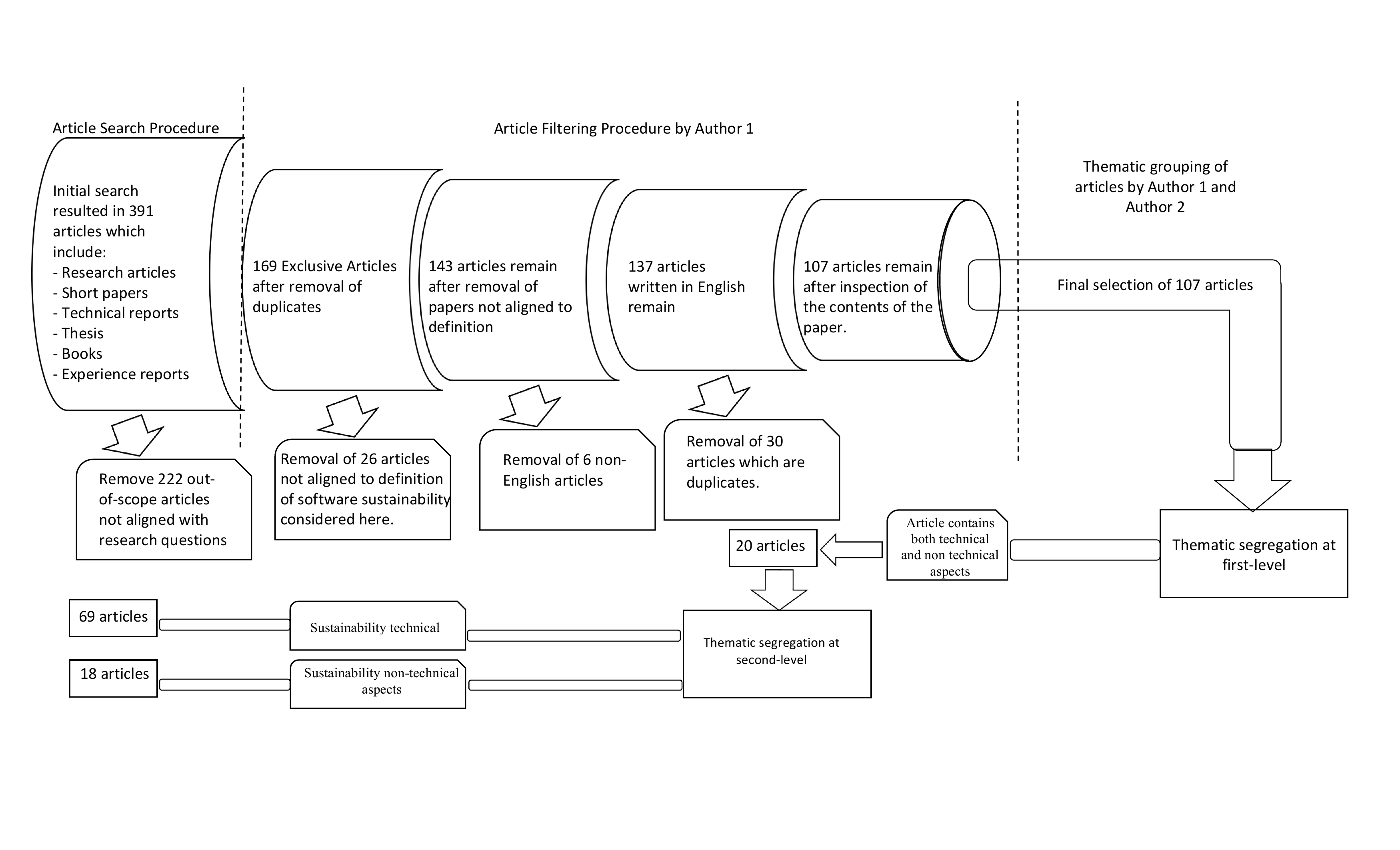}
  \caption{Methodology of the article filtering process of this review}
\label{fig3}
\end{figure*}

Many papers were found which talked about software sustainability, failures, etc with the words \textit{"software + sustainability"} in it. Initially, those were included since a specific definition of sustainability was not followed. As a result, the scope of the study became very wide. Adoption of the definition of software sustainability provided by the Software Sustainability Institute enabled narrowing the scope and addressing in-depth research questions. As a result, the above factual criteria were determined which added value by helping to filter out a number of articles. For example, the articles which deal with building software which adhere to environmental sustainability were beyond the scope of this paper, hence those were excluded. Such aspects can be included in future research. Initially, we removed duplicate articles together with the ones which are totally unrelated to the scope as shown in Figure \ref{fig3}. Following are the step-wise inclusion and exclusion criteria.

Inclusion criteria 1: Articles obtained by primary and advanced search with strings of words specified earlier.

Exclusion criteria 1: Exclude repeating articles. This included articles with the same results published by the same authors in different versions and venues.

Exclusion criteria 2: Exclude articles which are beyond the definition of sustainability of the Software Sustainability Institute.

Exclusion criteria 3: Articles not written in English Language are excluded.

\begin{figure}
  \centering
  \includegraphics[width=10cm,height=7cm]{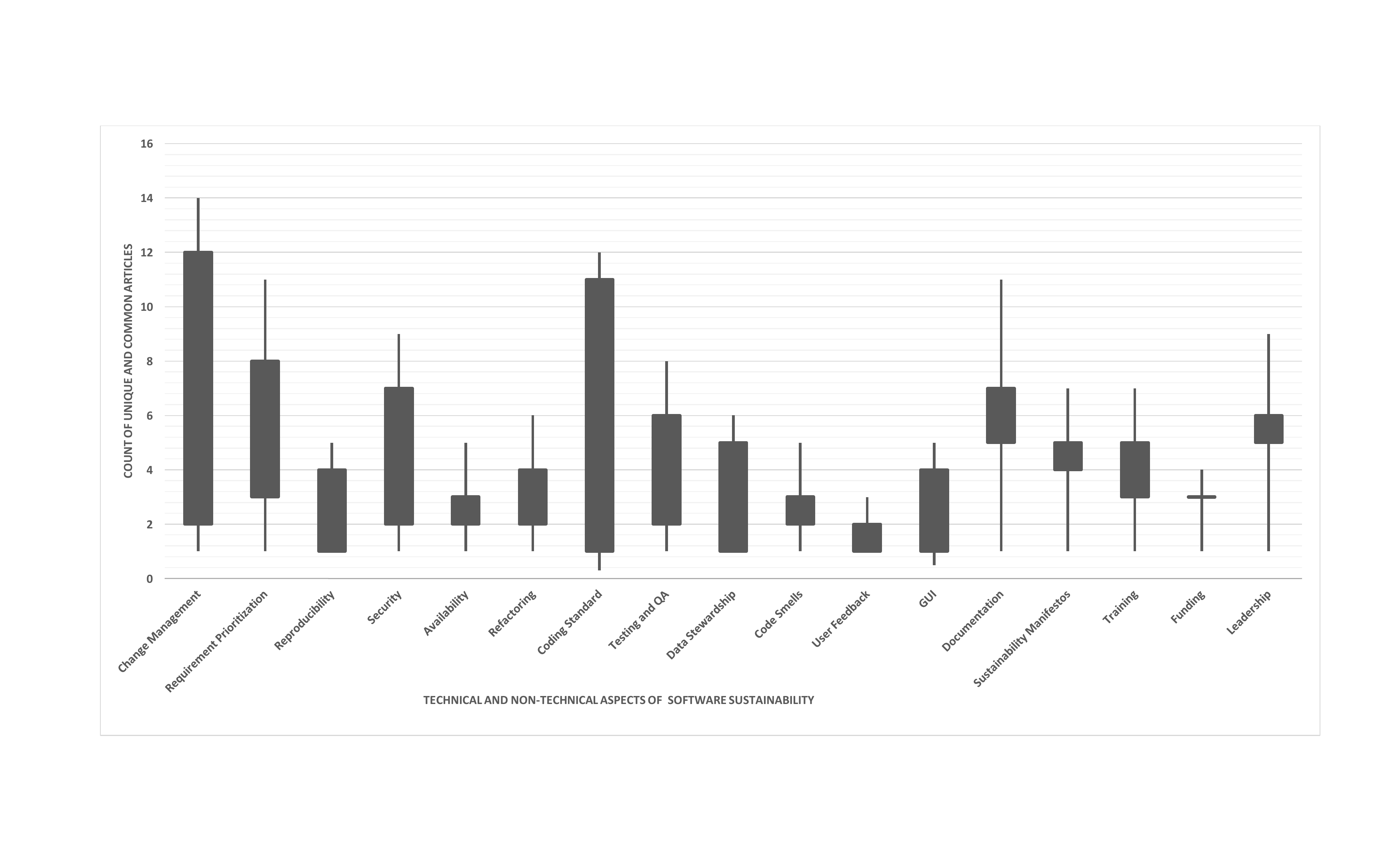}
  \caption{Selected articles belonging to different technical and non-technical aspects of software sustainability}
\label{technical}
\end{figure}

\begin{figure*}
  \centering
  \includegraphics[width=12cm,height=8.6cm]{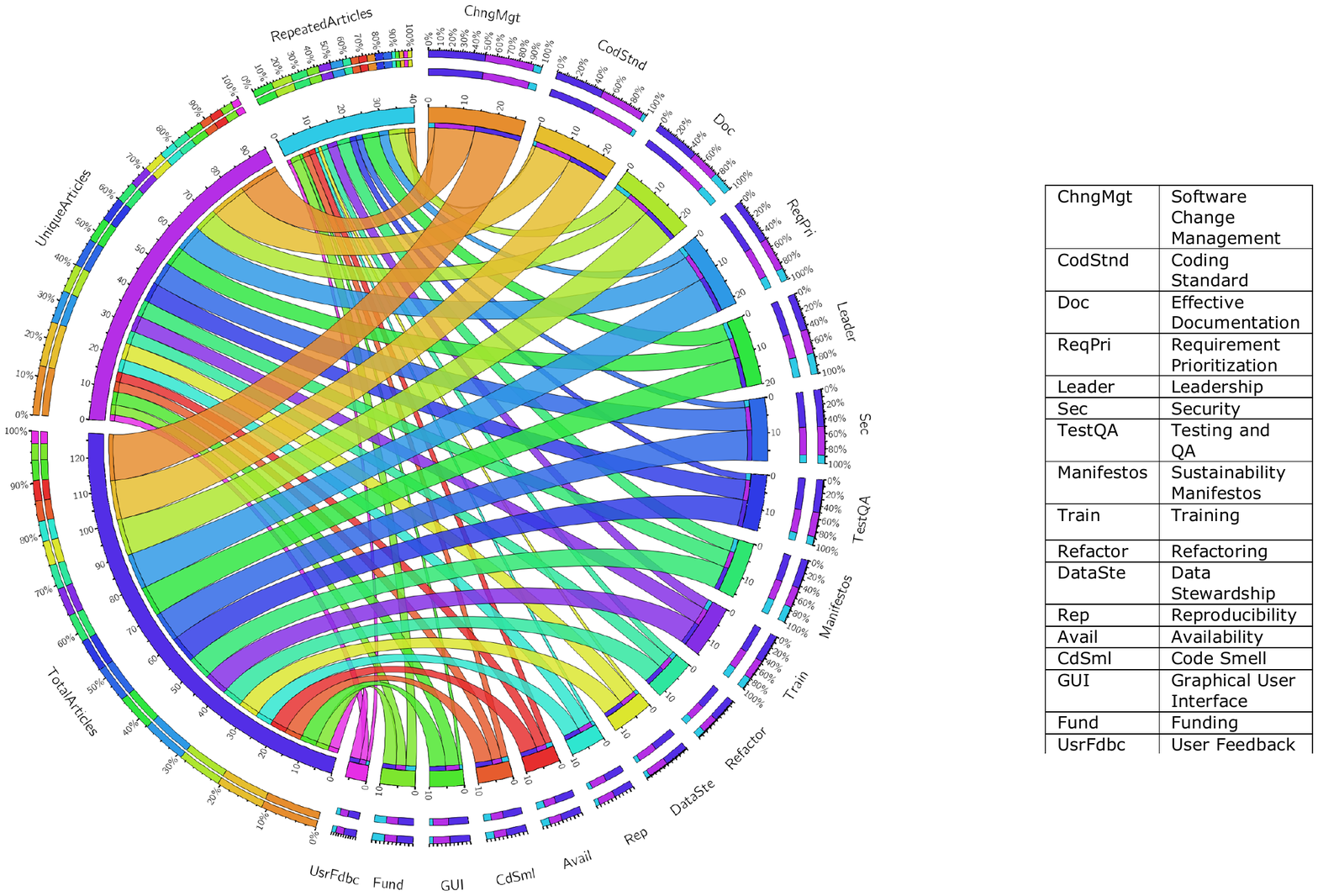}
  \caption{Graph showing causal relationship between various aspects of software sustainability.}
\label{commonpapers}
\end{figure*}

After applying the more detailed inclusion/exclusion criteria, we were left with 107 articles as irrelevant articles could be discarded. Full-texts of the selected articles were obtained and analyzed. Next, the form suggested by Kitchenham et al. was used for data extraction from the papers. Figure \ref{technical} provide the statistics related to the number of articles in various fields. It is seen in Figure \ref{technical} that the total number of articles exceeds 107. This is due to the fact that articles discussing both technical and non-technical aspects are shown multiple times in the figure. More specifically, there are 20 articles which contain information regarding both technical and non-technical aspects of software sustainability as shown earlier in Figure \ref{fig3}. Causal relationships between the various technical and non-technical aspects of software sustainability have been provided in Figure \ref{commonpapers}. The links inside the cylinder of the diagram show the relationships of different articles to the various software sustainability aspects.

\subsection{Validity of the review process}
Following the review protocol as stated by Kitchenham et al. \cite{kitchenham2013systematic}, the validity of the search and filtration processes could be justified. The review protocol was planned by first author and discussed with the second author, and all compliments and criticisms were noted. The protocol was re-modelled according to the feedback of the second author. For example, the second author suggested an advanced search together with the preliminary search to narrow down on the relevant research articles which was a useful feedback.

The first author of the paper was responsible for conducting initial searching of papers which introduced self-report bias.  The agreement/disagreement on the selection of final list of papers between the two authors was measured using the Cohen Kappa statistic, the initial value of the Kappa was found to be 0.46 which is considered as to moderate agreement. The disagreements were discussed and resolved. For example, after detailed discussion regarding exclusion of articles which were related to software aiding in environmental sustainability, both the authors reached an agreement of excluding those papers. The reason for this decision was influenced by the fact that the articles discussing software and environmental sustainability did not fall within the definition of software sustainability which was adopted in this study. 

To further eliminate the possibility of self-report bias, a test-retest approach was implemented \cite{kitchenham2013systematic}. This included generating random samples of size 20 from the primary list of articles. Then, 4 articles were randomly selected from the sample which were replaced with four other articles not in the sample. The later selection was also random. This ensured consistency of the sampling technique. The papers in the random samples were re-evaluated by both the authors after initial screening to verify the consistency of the review protocol.

\subsection{Thematic Synthesis}
\begin{figure*}
  \centering
  \includegraphics[width=14cm,height=7cm]{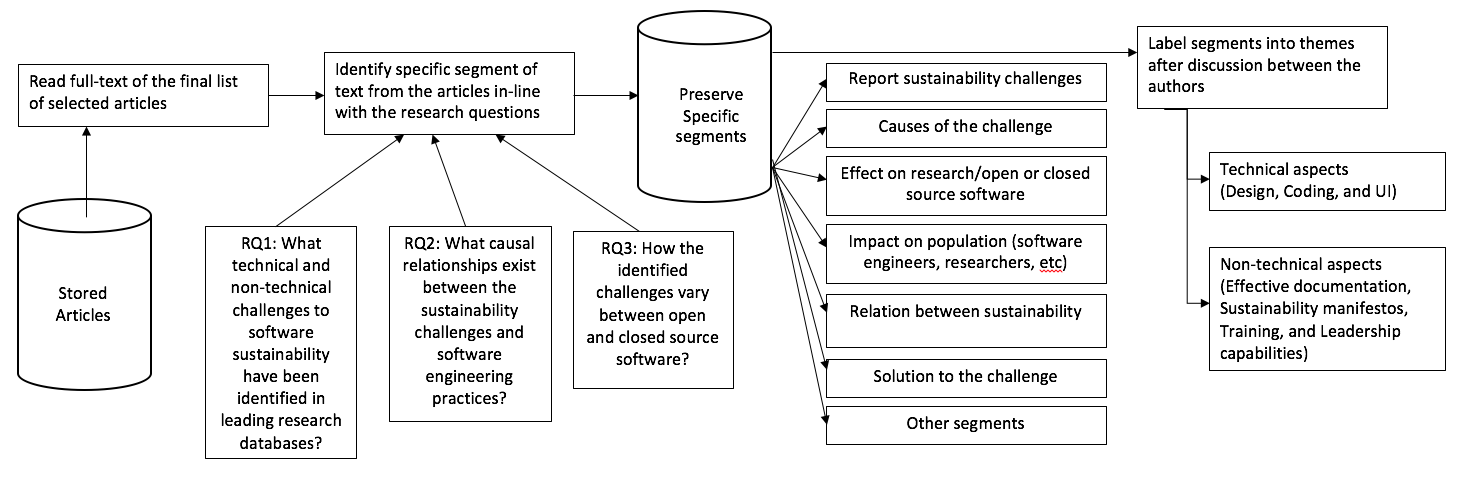}
  \caption{Thematic synthesis for the review process}
\label{fig4}
\end{figure*}

During initial screening of the articles, specific segments of the text were identified which contained answers to the research questions. Next, those were narrowed down to specific phrases and recorded in the data extraction forms. Overlapping amongst the labels needed to be removed, and this was achieved by discussion between the authors. We observed that technical aspects of software sustainability can be categorized based on the various stages of software engineering practices during SDLC. Hence, we proceeded to translating the labels to specific themes related to SDLC. Initially, we selected themes like software design, coding and user experience. Those themes were further granularized into more specific themes. Hence, the themes were not pre-selected, rather those were a result of evaluating papers on the basis of the research questions followed by narrowing down to specific sustainability aspects. The process of selecting the themes based on research questions is shown in Figure \ref{fig4}. 

As stated previously, our goal is to summarize the methodologies of software sustainability from both technical and non-technical viewpoints. For technical aspects, an \textit{approach based categorization} \cite{petersen2015guidelines} was done on the previous research papers. We determined from existing literature that system-wide sustainability can be achieved by integrating specific practices at various stages of the \textit{SDLC}. Those practices included both technical and non-technical factors. We categorized the current technical approaches to sustainability based on software design, programming principles, and user feedback. Based on the approach oriented categorization and snow-balling discussed in previous sub-sections, we identified the related research papers. 

We discussed the software design principles which ensure sustainability as well as the programming, refactoring, and user-experience practices that can help to make a software sustainable. Previous publications based on the aforementioned technical aspects of software design principles, programming, refactoring, and GUI have been described and compared in this review.

It was highlighted earlier in this review that non-technical aspects of software sustainability are equally important as the technical methodologies. We identified important non-technical attributes for software sustainability which are currently in place, compared and contrasted their roles in ensuring software sustainability. In the next section, we identify the research papers which state the importance of various components of software design towards sustainability.

\section{Software Design Principles}
\label{design}

This section identifies the panorama of related research articles regarding different software design aspects of sustainability which have been identified in the previous section. They are discussed here to greater detail. We identify the definitions of the sub-categories in this section. Next, we provide a tabulated comparison to show how the identified factors play a role to achieve sustainability of open source and closed source software. Afterward, we discuss how each paper plays a role in sustainability. Next, we describe the research efforts on how the mentioned factors can be achieved.

\begin{enumerate}

\item \textit{Software Change Management (SCM)} is the task of tracking and controlling changes in the software, which is part of cross-disciplinary field of configuration management. The steps in \textit{SCM} include revision control and establishment of baseline. They are significantly important for software sustainability since proper \textit{SCM} will ensure that the changes are integrated in a consistent manner, which will make sure that the software continues to evolve with changing user requirements. 

\item \textit{Software Requirement Prioritization (SRP)} is the process of determining which candidate requirements of a software product should be included in certain releases. \textit{SRP} is used to reduce the possibility of important requirements being ignored or delayed during the software development process. By making sure that the critical requirements are included all the time, \textit{SRP} will help to improve the sustainability of the software.

\item \textit{Software Reproducibility} is the capability of a software to repreduce results which are already generated using the same data. It is one of the main principles for building reliable software for scientific and commercial purposes. The increase in reproducibility of results will help users of scientific software run their tests a multitude of times and get the same results, thereby increasing the chances of such software to sustain over a significant time period. 

\item \textit{Security} in the context of software design includes the steps which are applied in defense to protect its processes, workflows, and data against any kind of intrusion and malicious attacks. Making a software secure will increase its acceptability among the users, hence increasing its longevity.

\item \textit{Software Availability} can be defined as the percentage of time for which the software will function without stoppage. Ensuring high availability for a certain time period will increase software reliability, therefore users will want to use it more as compared to other competitive software. Hence, the software can sustain longer due to increased user following.

\end{enumerate}
\begin{center}
\begin{table*}
\begin{tabular}{ |p{2.2cm}|p{7cm}|p{6cm}| }
\hline
\multicolumn{3}{|c|}{Design aspects of sustainability} \\
\hline
Sustainability Criteria& Open Source & Closed Source \\
\hline
  Change Management &Despite the availability of certain change management tools, using those for open source software is a challenge due to the lack of a well-defined change management process and team \cite{staahl2018dynamic}. Further research is needed to devise effective change management procedures for open source software. & Source code management and release management software are mostly client-server approach which are often built for specific operating systems. Examples include \textit{IBM Rational ClearCase}\cite{clearcase} and \textit{Endeavour}\cite{endeavour}.\\ 
\hline
Requirement Prioritization & The requirement specification documents are not properly maintained. Contributors find it difficult to identify high value software requirements which is a challenge to sustainability \cite{penzenstadler2014safety}.  & More structured mechanism of collection and prioritization of requirements. However, they are only available to the software owners and the users do not have access to them \cite{Chitchyan:2016:SDR:2889160.2889217}. \\
\hline
Reproducibility & Research is needed to develop platforms for sharing data and workflows more effectively. As data intensive scientific discoveries become common, reproducibility of scientific software becomes challenging \cite{kitzes2017practice}.  & Tools for provenance tracking for reproducing results and scientific workflows are available \cite{durdik2012sustainability}. However, those tools need to be procured before those can be used. \\
\hline
Security & The source codes are available to users who can download, modify and fix bugs. However, wide availability of source code makes it accessible to the hackers to practice with it \cite{penzenstadler2014safety}. Empirical methods like \textit{Onion Framework} \cite{imran2016web} need to be tested for sustainable security \cite{Lago:2015:FSP:2830674.2714560}.& Source code is mostly managed by the vendors and general public cannot access it, any change needs to be requested to the vendor \cite{penzenstadler2014safety}. This reduces that chances of hackers getting access to codes \cite{deal2004cisco}. \\
\hline
Availability & It is the percentage of time for which the software will function without stoppage \cite{Marwah:2010:QSI:1773394.1773405}. Easily accessible code allows software engineers to integrate certain features into the software which may not sustainable in long run \cite{Marwah:2010:QSI:1773394.1773405}. & Software is more restricted since the source code is inaccessible and not viewable by general users \cite{Marwah:2010:QSI:1773394.1773405}. However, such limitation contributes towards increased security and reliability. Licensing provides the right to others to use and modify the software's assets.  \\
\hline
\end{tabular}
\caption{Design attributes required for software sustainability}
\label{designtable1}
\end{table*}
\end{center}

Table \ref{designtable1} shows software design attributes which are core to ensuring sustainability from the software engineering perspective. It is seen that software change management is a sustainability requirement from the perspective of both commercial and research software. Change management is key to ensure that the software can evolve over time which is a fundamental aspect for achieving its sustainability \cite{yan2018ontology}. Requirement prioritization is another key attribute which must be present in all forms of software to achieve sustainability \cite{Chitchyan:2016:SDR:2889160.2889217}. A software may have many requirements initially, however a proper software development procedure needs to be followed which will ensure that the key requirements are addressed with a higher priority \cite{neela2017modeling}. 

For mission critical research software which have single or a handful of software requirements, all the requirements are functional and they are all of equal priority. For example, a research software used to simulate the state transition diagram of vehicles using finite automata have a single high priority requirement. In this case, there is no major reason to prioritize the tasks of the software. Any other cases which have multiple user requirements will need prioritization to identify the important ones from the list. 

Research software, both closed and open source, will have a large group of initial set of functional and non-functional requirements, however it is the responsibility of the software professionals to achieve the functional requirements before the non-functional requirements \cite{Chitchyan:2016:SDR:2889160.2889217}. Any functional requirement which could not be addressed in a version of the software needs to be clearly mentioned to the users \cite{Chitchyan:2016:SDR:2889160.2889217}. Hence it is observed that one of the important aspects of a software surviving for a long term is to identify and implement the requirements which are important to the user groups \cite{Ojameruaye:2016:SDP:2889160.2889218}. Different ways to prioritize user requirements are shown in \cite{yan2018ontology} and \cite{penzenstadler2014safety}.

As shown in Table~\ref{designtable1}, \textit{Reproducibility} is a sustainability attribute which will be required by all open source general purpose software. However, it is needed for a specific group of closed source software as well. \textit{Version controlling} is another important attribute which is needed to ensure sustainability of all types of general purpose software (open and closed sources) mainly to address the various versions of the same software which are released over a given period of time \cite{ver_2018}. It also helps to achieve co-ordination and synchronization of software development when a large group of software developers are involved. Security is another important attribute for almost all kinds of software to sustain over longer period of time \cite{penzenstadler2014safety}. 

It is seen that, many software nowadays are web-based systems which require secured sharing of user data such as credit card, date of birth, etc \cite{kelly2019data}. These software need to preserve privacy and integrity of user data. Hence, they need high levels of security as illustrated in \cite{imran2016web} with respect to software running on the cloud. Software which do not require user information or transfer any data may not need high levels of security \cite{nakagawa2018sustainability}. Software availability is an important feature of sustainability which makes sure that it is functioning most of the time \cite{hill2017integrated}. Even if the system needs to be shut down for maintenance, upgrades, etc, the users must be made aware of the issue well in advance. It provides reliability and trust of the user on the software itself. Hence, it is required for all types of closed and open source software.


The remainder of this section identifies the related literature of the stated categories of design aspects of software sustainability.

\subsection{Software Change Management}

Stahl and Bosch \cite{staahl2018dynamic} stated that sustainable software are the ones which allow small changes to be integrated in the smallest possible time. Scientific/research open source software with limited number of contributors lack this provision since they require longer time to integrate specific extensions, addition of new features, and tuning of existing functionality for better performance \cite{staahl2018dynamic}. To solve this issue, the authors chose to follow a Service Oriented Architecture (\textit{SOA}) to achieve better accommodation to changes within a short span of time \cite{lago2010service}. It is worthy of mentioning that web-based applications can use features like atomic design \cite{atomic} which is worthy of integrating changes and modification to the various components of the software. Use of atomic design for sustainable component modification is desirable, however the time cost to integrate such methods into real-life software design requires further exploration.

Prediction of time to address an issue has been an area of significant research interest \cite{da2014empirical}. After addressing an issue, it must be incorporated in one of the releases for the general users to use it. Research have shown that even though the issue has been addressed, integration in  a release often gets delayed due to load on the integration team and sudden surfacing of non-negligible changes to the software associated with integration. Anvik et al.~\cite{anvik2006should} investigated and concluded that although the issues have been solved well ahead of release time, un-predicted, important integration changes could not be determined well-ahead. This feature is expected to negatively contribute towards improving the sustainability.

The importance of addressing both the technical and non-technical approaches for software sustainability have been addressed by Winters~\cite{8443579} which clearly identified the need for non-atomic refactoring of software which were designed to last for decades and adapt to changes. 

Every scientific software development should be considered as a software project, and popular tools used in the industry to maintain projects and ensure their sustainability must be followed \cite{jimenez2016standing}. The tools will vary from project to project. Mostly, current software projects are maintained by using version controlling tools for change management or simply by commenting on the code. Although version controlling tools contribute partially to software sustainability, however, they do not cover all the aspects \cite{jimenez2016standing}. Other design approaches like reproducibility, requirement prioritization, security and availability should be considered as well.
 
US National Research Council first mentioned about the necessity for best practices to achieve sustainability of software in the report which was published in 2003 \cite{wicherts2011willingness}. Based on the practices in \cite{wicherts2011willingness}, important directives were identified by \cite{stodden2013best}. They identified standards of code re-use, citation standards for code and version controlling as key aspects of sustainability for software engineering. Although they focused primarily on software used in the area of natural sciences, it is equally important to ensure that the same aspects for research software in all dimensions. For version controlling of code, they suggested the use of tools such as \textit{Github} \cite{dabbish2012social} and \textit{Bitbucket} \cite{bitbucket}. Next, they requested to provide Github links for citing codes. 
 
 Researchers from \textit{ELIXIR} \cite{elixir} and \textit{The Carpentries} \cite{carpentries} have come together in Jiminz et al. \cite{mateusz_kuzak_2018_1420462} to present 4 best practices for developing open source software (\textit{4OSS}) preserving sustainability as well as maintaining the recommendations of \textit{4OSS} \cite{jimenez2017four}. They have identified that the software project should be made open source as soon as possible after conceiving as it will help in ensuring sustainable design. Proper procedures should be maintained using popular tools for change management. The longer a project stays in closed source, the harder it becomes to make it open source later. Discovery of the software must be made easier by making the metadata of the software available through a public registry. However, improving sustainability of software which cannot be made open source has not been addressed. Also, the financial challenges related to the sustainability of open source software were not addressed.

\subsection{Requirement Prioritization}

Design for precision is needed for scientific software, the importance of accuracy is manifold for sustenance \cite{kitchenham2004evidence}. Hence, prioritizing performance related requirements in a research software is important for it to sustain \cite{penzenstadler2014safety}. Precision variables like number of decimal places, execution time units, etc need to be programmed into the system as wanted by the users of the scientific software \cite{penzenstadler2014safety}. Minimalistic design catering to exceptional needs means that software designers must emphasize on the exceptional features which might be critical to scientific software which may not be that important for regular software \cite{penzenstadler2014safety}. For example, log collection will be an important feature for most of the scientific software. Also the software which are replacing the operational interface of scientific hardware will be needed to be designed as per the looks of the hardware itself \cite{morris2009some}. For example, the User Interface (UI) must have dials, knobs and meters where necessary to mimic the hardware look \cite{morris2009some} so as to present a user-friendly interface to the users with which they are adept. Exploring such techniques of scientific/research software to ensure sustainable operation requires further research.

Special emphasis has been provided to sustainable software design for scientific usage in the field of Computer Science \cite{katz2015report}. The importance of prioritizing specific requirements of scientific software to assure sustainability have been identified and guidelines have been provided in effective designing of such software. Backgrounds of problems were evaluated and based on them nine guidelines were provided which, if followed, will improve usability of scientific software. However, application of such guidelines to a real-life scientific software in order to evaluate their performance on increasing sustainability is an area of future research interest. 

Current principles and standards for sustainability design at requirements engineering phase have been highlighted by Chitchyan et al. \cite{Chitchyan:2016:SDR:2889160.2889217}. They have primarily focused on the challenges faced by software engineers to understand and integrate sustainability in their design as software increasingly gets included in the social and technical fabric of our society. However, case specific challenges of software sustainability and mitigations of those have not been addressed.

Contrastin et al. \cite{contrastin2016supporting} discussed about prioritizing lightweight specifications which play important role in maintaining software by ensuring all major and minor changes to software's initial specification adhere to certain sustainability properties. They took numerical software as a case study and provided two lightweight specifications for them, which are units-of-measure-types, which aim to specify the physical units of numerical entities in a program \cite{contrastin2016supporting}. Next, they mentioned stencil specifications which identify the pattern of data access used to access arrays of numbers. Inflexible software is difficult to sustain and does not reflect the viewpoints of the coders. Most of the times coders try to pass their mindset to the viewers of the code by writing comments \cite{contrastin2016supporting}. However, comments in the code need to be updated from time to time \cite{contrastin2016supporting}. They also mentioned that software specification languages find it difficult to illustrate the detailed mathematical models in a software in fine granularity. As a result, lightweight specification and verification is expected to provide an intermediate solution by establishing a relationship between implementation and model of software \cite{contrastin2016supporting}.

Earlier work highlight the time which is needed to prioritize requirements in  a software system \cite{anbalagan2009predicting}, \cite{giger2010predicting}. This information is critical for software project managers for allocating resources. Prioritization of issues is also important to solve them in a sustainable manner. These technologies need to be tested on real-life software and their performance need to be evaluated.

\subsection{Reproducibility}

Over the last few years software engineers have provided great emphasis on software reproducibility. They have tried to achieve this by popular techniques such as containerization, documentation, orchestration, dependency management and testing \cite{kitzes2017practice}. The goal is to provide a generic pipeline to abstract the workflow, experiments, simulations and analysis. Ensuring reproducibility in research is significantly important to making research software more sustainable over time.

The time wasted by researchers in reproducing previous research have been highlighted by Jimenez et al. \cite{jimenez2017popper}. It was stated that an effective mechanism is needed which will enable the researchers to reproduce previous research results without much effort \cite{jimenez2017popper}. How this can be achieved is an area of significant research interest of sustainable research software engineering. Reproducibility encourages users to use the software more often, thus increasing its chances of remaining usable for a long period of time \cite{jimenez2017popper}. There needs to be an effective way via which the scientific community can reduce the time spent on re-implementing workflows, data and model \cite{jimenez2017popper}. A client \textit{(CLI)} based approach which can be used by researchers to reproduce previously published scholarly works is an area of research interest.

A software called \textit{CodeOcean} is described by Staubitz et al. \cite{staubitz2016codeocean} which provides a practical programming exercise to run software codes and evaluate reproducibility of research experiments. The cloud based software requires users to set up accounts, submit their codes, write CLI scripts specifying file paths and datasets before finally launching the runs. This software is currently being used by organizations like IEEE \cite{oceancode} to test the reproducibility of results in various research papers. More diversified and widespread use of such software will ensure sustainability of research applications. However, detailed research efforts are needed to make tools like \textit{CodeOcean} language independent so that they can check research results generated in any language and thereby prove sustenance.

The heterogeneous design procedure of research software and the culture of isolated development of those lead to the lack of reproducibility \cite{crouch2014software}. Implementing software engineering practices to design and building sustainable research software which has the capability of reproducing its results has limited or no reward \cite{crouch2014software}. This non-rewarding factor leads to developers finding quick fixes to the problems which arise in research software without keeping reproducibility in mind \cite{crouch2014software}. In the world where hardware is becoming obsolete in every couple of years and software becoming a part of capital expenditure in almost all research institutions, there needs to be advocacy for following effective design guidelines to create reproducible software \cite{crouch2014software}. As a result, there are requirements for increased focus on promoting training on producing reproducible scientific/research software \cite{crouch2014software}. This will ensure research software meet same level of reproducibility standards and in this way become more useful to researchers, hence becoming more sustainable.

\subsection{Security}

Software security is the idea of designing and developing software so that it continues to function as normal under malicious attacks. Penzenstadler et al. \cite{penzenstadler2014safety} labelled software security as one of the primary requirements of the 21\ts{st} century when software started to take over as ubiquitous services for most of the daily functions of life. In the 21\ts{st} century, the software are designed to support almost all of the major industries \cite{penzenstadler2014safety}. Many of these software need to handle complex work procedures of the industry \cite{penzenstadler2014safety}. The flaws in security design of those reduce sustainability. Hence the goal of the 21\ts{st} century is to build sustainable software via ensuring security, as stated in \cite{penzenstadler2014safety}. 

Lago et al. \cite{Lago:2015:FSP:2830674.2714560} identified security as a traditional software quality attribute. Norris \cite{norris2004mission} stated that one of the goals to achieve sustainability of mission critical open source software is to publish all the security loopholes and try to engage the community to solve those. It was stated that via proper communication of security vulnerabilities, patches for 90\% of the reported issues became available within few days \cite{norris2004mission}. This case study was based on a mission critical open source software supported by \textit{Cisco \cite{deal2004cisco}, GNU \cite{anderson2007towards}, FreeBSD \cite{jorgensen2001putting}} and \textit{Ripe} before \textit{CERT} \cite{allen2001cert}. A paper from NASA stated that as long as the security specification of national security are preserved, the software will sustain over a long period of time in the USA and the rest of the world \cite{moran2003developing}. 

A literature review for combining various software stretching across geographical boundaries which work together to serve a common purpose (i.e. System of Systems (SoS)) that no single software can achieve have been identified by Guessi et al. \cite{Guessi:2015:SLR:2695664.2695795}. Although there are some studies which support secured development of SoS \cite{6034903}, \cite{dagli2008system}, there is a lack of research effort on how to ensure the sustainability of such software via achieving security. In this context, they presented a panorama of the various architectures which can be used to design and develop secured \textit{SoS} \cite{nakagawa2013state}, however a study on how they can be made sustainable is needed to be considered to a detailed extent.

\subsection{Availability}

Marwah et al. \cite{Marwah:2010:QSI:1773394.1773405} stated that recently sustainability has been a critical point to consider for software. Since availability is a critical aspect of sustainability, it needs to be addressed in this literature. Software availability is measured by the resilience of the software program together with the hardware platform (e.g., Data center) which hosts the software \cite{Marwah:2010:QSI:1773394.1773405}.  Marwah et al. computed availability by \textit{Stochastic Petri Net (SPN)} models while an energy-based Life-Cycle Assessment (LCA) approach was used for quantification \cite{Marwah:2010:QSI:1773394.1773405}. They showed that the method works on real-life data center architectures. By analyzing five different architectures, it was concluded that quantification of sustainability provided important inputs towards deciding which architecture will be best suited for hosting specific software to achieve sustainability.

A software availability model considering the number of restoration actions have been proposed by Tokuno and Yamada \cite{tokuno2003relationship}. They stated that when there is an imperfect bug fixing environment and practice within the team, an effective software availability measurement scheme should be deployed to ensure that the software continues to perform for a long time period, i.e., it sustains \cite{tokuno2003relationship}. They used Markov process to estimate the alternating behavior between up and down of time-dependent states of a software \cite{tokuno2003relationship}. The lesser the number of restoration, more reliable the software. However, to what extent this availability measurement scheme improves sustainability was not addressed.

Service Oriented Architecture (SOA) is designed thinking 24X7 availability of services. Lago et al.~\cite{lago2010service} contradicted that ensuring technical sustainability via 24X7 availability may threaten environmental sustainability of the software since it will lead to increased carbon emmission. They argued that if we keep the software and the underlying hardware running 24X7 to achieve 100\% availability, it will emit a lot of carbon and heat even during the off-peak hours when the software is not used. As a result, carbon emission will increase. It is expected that under current circumstances, carbon emission is likely to increase to 75\% by the end of 2020 \cite{lago2010service} instead of the targeted 20\%. Although the fact mentioned by the researchers is important, it lacks empirical investigation on how to assure 100\% availability at reduced carbon emissions for mission critical software.  

\section{Software Coding Principles}
\label{coding}
This section identifies the related literature which show why the coding principles of software play an important role in ensuring sustainability. 
Broadly, the coding principles of software for sustainability can be divided into five sub-categories. They are rigorous \textit{Refactoring, Following a Coding Standard, Testing, Data Stewardship,} and \textit{Code-Smell Detection}.

\begin{enumerate}
    \item \textit{Refactoring} software is the process of changing the object-oriented structure in such a way that it does not interfere with external functionality of the software \cite{1265817}.
    \item \textit{Coding Standards} are the conventions of software programming which make the code readable and easy to maintain. It is one of the desirable properties since a software which sustains for long periods of time is likely to undergo many changes in the code, therefore a coding convention will make the code readily understandable so that changes can be integrated easily.
    \item \textit{Testing \& Quality Assurance (QA)} is the process of inspecting and assuring the quality of the processes used to develop the software. It includes the entire process of verification, validation and testing.
    \item \textit{Data Stewardship} is the next attribute for software sustainability. It is the framework which supports generation, management, storage, transfer and archiving of big, medium, and small data volume together with metadata \cite{matsubayashi2017conceptual}. It is useful to achieve data consistency across all users and platforms in a software, which is a critical attribute of survivability for all kinds of software, be it closed or open source.
    \item \textit{Code Smell} is the aspect of software code the detection of which indicates a deeper problem. A code smell at a particular place in a code indicates some changes are required in some other locations of the software as well. 
\end{enumerate}

\begin{table*}
\begin{tabular}{ |p{2cm}|p{5.5cm}|p{5.4cm}|  }
\hline
\multicolumn{3}{|c|}{Coding Aspects of Sustainability} \\
\hline
Sustainability Criteria& Open Source & Closed Source \\
\hline
Refactoring & Lack of full-proof and sustainable code refactoring mechanism leads to software decay over time. On the other hand, defects may be detected and refactored faster if there are a large number of responsible contributors. &Continuous source of funding enables investment of resources for code analysis and refactoring, thereby achieving technical sustainability. \\
\hline
Following Coding Convention &Contributors belong to diverse backgrounds, hence it may become difficult to follow a single coding guideline \cite{8500171}. Some examples of coding standards include Linux kernel coding, GNOME programming, GNU programming standard, etc \cite{8500171}. & Generally a specific coding standard is followed since the software are designed and maintained by engineers of a particular organization, group, or company \cite{8500171}. Example is CERT coding standard which provides a outline of secure coding for commercial software \cite{8500171}. \\
\hline
Data Stewardship &Although many open source tools like \textit{CUBRID, Knowage, BIRT,} etc are available, which tool will be sustainable for what type of data requires further research \cite{matsubayashi2017conceptual}. & Primarily commercial data stewardship software are used \cite{matsubayashi2017conceptual}. Examples include \textit{Oracle Business Intelligence, IBM Congross, SAP Netweaver}, etc. \\
\hline
Testing and Quality Assurance    &Lack of a dedicated test team and lesser accountability may lead to a number of components  to remain untested. & Commercial software focuses on unit testing as well as vulnerability assessment \cite{staahl2018dynamic}. They also use extensive documentation in testing \cite{staahl2018dynamic}.\\
\hline
Code Smell Detection & Lack of funding and established smell detection practices prevent code smells to be detected and refactored in many open source software \cite{Ojameruaye:2016:SDP:2889160.2889218}. Those smells exist in the software and lead to technical debt \cite{Ojameruaye:2016:SDP:2889160.2889218}. & Commercial software employ code smell detection and solution tools targeting variety of code smells (primarily top 22 smells) and using lightweight approaches. \\
\hline
\end{tabular}
\caption{Coding principles required for software sustainability.}
\label{CodingTable}
\end{table*}

Table \ref{CodingTable} summarizes the important coding principles required to make software sustainable. Similar to Table \ref{designtable1}, the comparison is established between open and closed source applications. It is seen in the table that regular, rigorous refactoring of software code is a priority when it comes to sustainability and longevity of the software. Refactoring is applicable for both open source and closed source software. 
Code smell detection is considered to be an important sustainability attribute of both open source and closed source software as well \cite{7337457}. It is the automatic and early detection of software code which cause deterioration of software functionality \cite{6385092}. It is required to ensure that the software quality is preserved through systematic and automatic code inspection. Penzenstadler~\cite{1223642} highlighted the need for continuous Testing and Quality Assurance (QA) practice from an early stage of software development lifecycle, highlighting the role of inspection in software testing and QA which is critical for a software to sustain and perform under changing business and functional environments \cite{8445538}. 

Although the results presented in \cite{8445538} are applicable for testing the sustainability of a robotic operating system, testing it on other environments is required so that the mentioned procedure can be generalized towards other kinds of software. Hence we identify it as an integral criteria of software sustainability for both open and closed source software. Carter et al.~\cite{8500171}  and Sanchez-Gordon et al.~\cite{sanchez2016understanding} emphasized the need for following a standard coding convention from an early stage of software development life cycle to ensure that the software code, data, and operating platforms can be managed as they evolve over a long period of time. As the software evolves, it undergoes many changes and maintenance becomes a challenge for both open and closed source software if a specific programming convention is not followed, thus hampering sustainability \cite{sanchez2016understanding}.

In the following subsections, we identify and discuss the sub categories of sustainability with regard to coding principles.

\subsection{Refactoring}

Winters \cite{8443579}  stated that software engineering is a study which has programming integrated with it. In addition, it has a detailed vision of ensuring the proper coding and refactoring standards so that software can efficiently survive over a decade \cite{8443579}. The paper provides an important case for non-atomic refactoring and improvement of standardizing C++ codes to achieve sustainability \cite{8443579}. However, how to use the non-atomic refactoring methodology in real-life software case has been addressed to a limited extent.

The importance of studying the experiences of stewards who developed and used a wide range of open source software tools to identify and focus on sustainability issues have been addressed in \cite{hdf.2018}. The \textit{Hierarchical Data Format (HDF)} \cite{hdf.2018} group has been working with the research community for 30 years, building open source tools to provide platforms for sustainable data storage, access and analysis. It has analyzed \textit{Github} and found that nearly one thousand repositories are based on open source codes \cite{hdf.2018}. Also, several broadly successful open source systems are centered around \textit{HDF} \cite{hdf.2018}. The experience of $PyTables$ \cite{hdf.2018} was shared which stated that whenever an existing code is reused or re-factored, there is a high probability that it will present unforeseen issues which require correction and addressing, thus reducing the time to market.

The use of \textit{Iterator} pattern in code-refactoring for increasing sustenance of a software called FLASH which is used in research of physics have been highlighted by O'neal et al. \cite{o'neal_weide_dubey_2018}. The existence of a test-suite was found to be highly important for effective implementation of the test process \cite{o'neal_weide_dubey_2018}. It was also mentioned that all complex research software need a test base before going into total implementation \cite{o'neal_weide_dubey_2018}. They suggested refactoring as a worthwhile experience to improve software sustainability. They stated the use of an iterator to implement a parallel version of a number of sequential processes in \textit{AMReX} \cite{amrex} of FLASH provided more confidence on the re-factored code. The cost-benefit trade-off was that extensive code needed to be done on the older versions whereas the ultimate aim was to replace entire older version of the code \cite{o'neal_weide_dubey_2018}. Use of encapsulation and modularity made the re-factored code clean as claimed in the paper. However, quantitatively analyzing the improvement on performance of sustainability have been addressed to a limited extent.

\subsection{Following Coding Convention}

People today are reliant on web services which can harness the power of the cloud \cite{petcu2011towards}. They want a single platform through which they can access all their resources. For example, a web application which helps them book tickets from almost all airlines in the world is certainly desirable. Also, sites like \textit{Airbnb} \cite{zervas2017rise} are using the Google calendar to insert booking into the renter's calendar and send them email notifications periodically. Web-based Application Programming Interface (API) allows the users to use services provided by a third-party directly into their applications \cite{petcu2011towards}. This aspect requires the web service developer to integrate the third-party API into their web application so that the users can use it \cite{petcu2011towards}. More specifically, this task of integrating the APIs becomes more challenging as their use become more complex \cite{petcu2011towards}. Even simple uses like data transfer using the API of third-party data storage services poses challenges \cite{petcu2011towards}. For example, retrieving data from \textit{Twitter} will require a single parameter passed to the \textit{GET} request \cite{petcu2011towards}. On the other hand, merging various branches of a software into \textit{Github} using their API present a bigger challenge. Petcu et al. \cite{petcu2011towards} emphasized the need to follow a proper coding standard to ensure smooth and sustainable API integration into the cloud. However, implementation of the suggested methods to a real-life application to test for improvement in sustainability have been identified to a limited extent.

Definition of software sustainability and relationship with clean coding practices have been provided by Aldabjan et al. \cite{aldabjan2016should}. An empirical mechanism to determine the time of sustainability in terms of days was proposed. They evaluated 3000 projects in \textit{Github} to identify relationship between sustainability and coding practices \cite{aldabjan2016should}. They stated that controllable lines of code, method induction, etc are acceptable metrices of good coding practices that result in ensuring sustainability \cite{aldabjan2016should}. However, they evaluated a wide variety of codes and did not specify on specific use cases, so the results are more generalized.

 Winters \cite{8443579} addressed that the primary challenge is to address the stability of a software in an era where user codes are constantly changing, APIs are getting upgraded and infrastructure libraries are themselves unstable, which in return pose a threat to sustainability. The growing complexity of dependency graphs is a critical issue as software grows over time \cite{8443579}. Hence, an industry standard coding convention which preserves software sustainability needs to be adopted \cite{8443579}. With growing complexity comes diamond graphs \cite{haughton2006review} which are multi-node cycles in a graph. Indirect users are found to suffer the most from diamond dependencies \cite{haughton2006review}. If there is a break, it is highly unlikely that indirect users will come to know about the cause of the break to be an API or third-party component \cite{haughton2006review}. The authors identified three strategies to ensure stable integration of third-party APIs primarily derived from the principles stated in \cite{hanssen2012longitudinal}. They are:
\begin{enumerate}
    \item \textbf{No change} \cite{haughton2006review}: The API does not change over the years. If the API provider confirms no change when the API is first integrated then it is desirable. However, the aforementioned case is highly unlikely in the world of modern programming languages. Anything involving complex data, modern networking concepts, or programming languages will not conform to such change.
    \item \textbf{Release all dependencies in a single entity} \cite{haughton2006review}: This approach is similar to a Linux release \cite{kroahhartman2008linux}. Small pieces of upgrades can be introduced in between releases which can be integrated. However, most of the users will not integrate new updates between releases thinking that it may jeopardize the sustainability of their software.
    \item \textbf{Live at head} \cite{haughton2006review}: This mechanism encourages stakeholders to use the most recent third-party dependencies. All the new upgrades are released based on the most recent platform. Any user who is using older versions of APIs will likely face challenges and compatibility issues with new platform.
    
\end{enumerate}

However, implementation of the suggested strategies to a real-life software to evaluate the improvement of sustainability was not provided.

Aue et al. \cite{aue2018exploratory} identified a list of general cases for majority of failures for integration of third-party web API into a web-based application following Service Oriented Architecture (SoA) \cite{aue2018exploratory}. The authors acknowledged the fact that web-based APIs are gaining popularity everyday, as a result, software applications are using a myriad of web APIs to achieve their objectives \cite{aue2018exploratory}. 

Bachmann et al. \cite{bachmann2000volume} mentioned that software components vary from each other with regard to their architecture, design, and implementation, hence the errors associated with the component integrations vary from case to case.  Citing the fact that no research has been conducted to group the errors for software component's under general umbrellas given the reality that the errors vary in nature from software to software, this paper aims to group them based on popular user feedback and they have come up with a grouping of eleven categories \cite{bachmann2000volume}. Additionally, they have aimed to identify where the groups of web API integration errors impact the most, again based on feedback from users \cite{bachmann2000volume}. Finally, they have identified some of the current industry practices which are followed by software developers to address those issues \cite{bachmann2000volume}. Hence the need to follow a standard coding practice in software development is highlighted. However, how standard coding practices play a role to improve software sustainability was discussed to a limited extent.

With the acceleration in acceptance of \textit{SOA} there is an increasing demand for web service providers to use third-party services with the broader goal of achieving their own objectives \cite{bormann2008third}. The increase in the number of general users of third-party components, demand for more web service integration by consumers is on the rise \cite{bormann2008third}. Hence, addressing this issue is definitely beneficial for the people as it will help them address an important bottleneck in the world of \textit{SOA} \cite{bormann2008third}. On the contrary, how the proposed solution for service integration applies to scenario-based applications with an aim to improve their sustainability requires further research.

For the purpose of terms of software re-use, the license, and \textit{Service Level Agreements (SLA)} need to be correctly drafted and implemented \cite{wilson2017good}. Wilson et al. \cite{wilson2017good} mentioned certain aspects of sustainable and re-usable code design taking note from the aforementioned publication which include commenting at the beginning of the code to explain what the code actually aims to achieve \cite{wilson2017good}. Second, the code needs to follow a standard convention to functions in order to make it clearer to understand \cite{wilson2017good}. Third, elegant use of existing libraries and variables will ensure that the software engineer does not re-invent the wheel \cite{wilson2017good}. Fourth, giving meaningful names to the functions and proper documentation will help other engineers and coders understand the code better \cite{wilson2017good}. Next, the authors emphasized on the creation of documentations for requirements of the project \cite{wilson2017good}. Afterward, importance was shed on providing a sample data set for the users to run and test the software program. This was followed by highlighting the importance of if-else statements instead of commenting to show the conditional statements. Finally, storing or archiving code in dependable research repositories was mentioned. Although the proposal is state-of-the-art, application of the above framework for coding convention on a real-life distributed software preserving sustainability needs to be conducted.

Bosch \cite{Bosch:2010:ACS:1842752.1842776} focused that most of the research efforts on third-party components are concerned with effective API design and handling common errors which occur in API integrations. However, no research has been done to address the need to group the errors under a standard convention so that they can be attacked and resolved holistically using a code management technique \cite{Bosch:2010:ACS:1842752.1842776}. This information is critical towards improving the API documentation and it will also provide knowledge of the common integration pitfalls \cite{Bosch:2010:ACS:1842752.1842776}. The use of \textit{Github} and \textit{Bitbucket} \cite{bitbucket} for error recording needs to be explored to a greater detail with respect to the sustainability aspects of software. Additionally, this paper also analyzes the potential impact of the errors on the users of third-party services, addressing the significance.

\subsection{Data Stewardship}

Citing the example that a large number of database services are moving to cloud platforms like Amazon Web Services (AWS), Abadi \cite{abadi2009data} identified a list of features which a database is required to have to give sustainable service. Those sustainable features include efficiency, fault tolerance, capability to run in heterogeneous environments, ability to operate encrypted data and ability to incorporate with business intelligence software \cite{abadi2009data}. Examples of \textit{MapReduce} and \textit{Shared Nothing Parallel Database} were highlighted \cite{abadi2009data}. Next, it was stated that efficiency of \textit{MapReduce} on parallel data is a matter of debate whereas parallel databases are designed to operate with homogeneous data types \cite{abadi2009data}. However, the paper stated the need for a hybrid solution for the cloud. Additionally, they highlighted that current open source databases do not possess the features in the list, so they cannot be sustained in AWS \cite{abadi2009data} in the long run. Finally they specified the need to design and develop a database system specifically for cloud which will achieve sustainable data management \cite{abadi2009data}.

The strength of \textit{Google BigTable} \cite{Chang:2008:BDS:1365815.1365816} for sustainable heterogeneous data management is noteworthy. It is used to store data from large applications like \textit{Google Index, Google Earth} and \textit{Google Finance}  \cite{Chang:2008:BDS:1365815.1365816}. It is optimized to store metadata in smaller sized files, thereby allowing storage of meteadata for large files. At the same time, it has the capacity to steward petabytes of data \cite{Chang:2008:BDS:1365815.1365816}. However, one of the limitations of \textbf{Google BigTable} is that the data throughput is dependent on \textit{Google File System (GFS)} \cite{Chang:2008:BDS:1365815.1365816}. Given the unusual interface of \textit{Google BigTable}, it is debatable as how many users have realistically been able to use it, particularly if they are accustomed to using relational databases.

Traditional data management systems implement internal consistency constraints within the software \cite{rajasekar2006prototype}. If these constraints need to be changed, it requires re-writing certain code blocks of entire software, which if done incorrectly will effect sustainability \cite{rajasekar2006prototype}. The experience of building the storage broker data grid was evaluated to identify the sustainability limitations of current database systems \cite{rajasekar2006prototype}. The authors proposed a rule based middleware for storage virtualization. However, the performance of such a middleware to ensure sustainability of real-life distributed data stewarding software needs to be evaluated to a greater extent.

Tracking user activity for usability via logging in \textit{MySQL} \cite{6982633} environment to manage data is important to track activities of users and provide them an easier mechanism to re-execute the commands which they run frequently. In fact, the reason for the astounding popularity of \textit{CLI}'s is because it allows the users to go through and re-execute the previous commands using the up arrow key \cite{6982633}. Also, logging user commands and storing them in a temporary history file can also provide the easy user access to the previously executed commands in case they are needed afterwards  \cite{6982633}. However, exploring such methods to improve sustainability of \textit{NoSQL} based applications was not addressed.

Cruz et al.~\cite{cruz_maria_j_2018_1419085} studied the effect of data stewardship on making software sustainable and reproducible. 
The authors identified that the key challenge from the perspective of data stewardship is to ensure that the basic level of service which is required by research software engineers are provided. Implementing the sustainability improvement ideas presented in the paper to real-life use cases is an area of future research interest.

\subsection{Testing \& Quality Assurance}

Robillard \cite{robillard2016sustainable} identified that although software design is key to its success, the points mentioned in design documents at an early phase of \textit{SDLC} evaporate over time as the software evolves, causing loss of artifacts and fading of developer's knowledge \cite{robillard2016sustainable}. To ensure sustainment of functionalities over time, there is a significant need to test the software exhaustively \cite{robillard2016sustainable}. In many cases, the testing approach is continuous. There can be various mechanisms in which initially specified designs might be preserved. Firstly, it can be formally maintained in  a design document \cite{robillard2016sustainable}. Secondly, it can be mentioned in email and medial conversations, chat groups among developers and project managers, etc \cite{robillard2016sustainable}. In both cases, designs are prone to evaporation. Hence, to ensure that sustainability design constraints do not evaporate they need to be tested regularly. 

Formally defined designs can evaporate as design drift and erosion occurs which significantly affects the sustainability \cite{tsantalis2006design}. This happens when the initial concept and design idea drifts from time to time as the project gets implemented \cite{van2002domain}. Lack of this practice causes design drift in many open source software. The importance of verifying and validating the various steps so that the drift from initial requirement is minimum should be explored. On the other hand, poorly maintained design information in emails and chat groups get lost, deleted, or discarded and become difficult to retrieve \cite{petre2013software}. As a design attribute, there is a trade-off between sustainability with other quality attributes. For example, to improve quality of software, we can add extensiblity features, however those features will introduce new methods which need to be documented, validated, and tested, which if not done properly will interfere with the sustainability of the software \cite{petre2013software}. By drawing examples from \textit{JetUML} \cite{jetuml} which is an open source \textit{Unified Modelling Language (UML)}, an illustration of various forms of sustainability was provided. Although it was stated that testing the software can assure its sustainment, which approaches improve the sustainability test methods were not identified.

Provisioning and testing for \textit{Domain Specific Language (DSL)} like \textit{XML}, \textit{UML}, etc, which are languages designed for a specific application domain, can increase sustainability of software \cite{van2002domain}. For \textit{DSL}, configuration files are needed to test and validate certain scientific applications which require large number of parameters to be passed before execution \cite{KOSAR2008390}. Hence, there should be a provision of configuration file which will contain all the parameters. The configuration file, when passed will be automatically read by the software before its execution \cite{KOSAR2008390}. The \textit{DSL} is more advanced and feature-rich compared to the configuration file \cite{van2002domain}. It allows users to programmatically define new functionalities in the software. It also allows the users to interact with the code at runtime, thus allowing them to modify, debug, trouble shoot and visualize the output. Hence, this configuration file can be used to test the outputs of the various software processes for which it has been designed \cite{van2002domain}. Although this is proven to improve sustainability, how \textit{DSL} can be used for testing the sustainability of scientific software have been addressed to a limited extent.

Achieving software sustainment through proper testing and quality assurance of software artifacts is an important practice \cite{seacord2003measuring}. The software artifacts such as source code, architecture documentation, and other architectural representations are necessary but insufficient to assess overall sustainability \cite{seacord2003measuring}. It is mentioned that completeness, consistency, and understandability will enable a software to become sustainable. Since these activities happen over the entire \textit{SDLC}, those require vigorous testing and quality assurance \cite{seacord2003measuring}.

\subsection{Code Smell Detection}

\textit{Textual Analysis for Code Smell Detection (TACO)} \cite{Palomba:2015:TAC:2819009.2819162} has been provided to make software more sustainable by using textual data to identify the type of code smell detected. The research is based on the argument that currently the focus on code smell identification is a structural process. The performance of textual approach was explored here with and accuracy of 67\% was obtained. By using textual approach, Polomba et.al \cite{Palomba:2015:TAC:2819009.2819162} was able to detect code smell which could not be differentiated via their structures. As a result, they made software suffering from those types of code smells more sustainable by improving detection of smelly code which ultimately set up its removal.

Azeem et al. \cite{azeem2019machine} provided a literature survey of the current machine learning techniques in code smell detection. They argued that the current machine learning techniques are not significant to detect all code smells and more work is needed \cite{azeem2019machine}. Their careful illustration of present research provides useful data on how machine learning techniques are used for detecting code smell which can then be removed, improving sustainability \cite{azeem2019machine}. However, this study is limited to 15 relevant papers in this area, which illustrates the need to carry out a more detailed study.

The scenario-based architecture for detecting code smells in specific scenarios are identified by Koziolek \cite{Koziolek:2011:SES:2000259.2000263}. The paper argued that although scenario-based architectures can detect smelly code and improve sustainability over the software lifecycle, those architectures are rarely implemented on real-life systems. Also, these scenarios are not included in the architecture level metrics, thus reducing their capabilities \cite{Koziolek:2011:SES:2000259.2000263}. It was stated that the smells which affect sustainability should be identified at an early stage of \textit{SDLC}, and no single architecture is able to characterize all the code smells \cite{Koziolek:2011:SES:2000259.2000263}. Various architectures should be combined at the process level and class level and should be monitored as the system evolves. \cite{Koziolek:2011:SES:2000259.2000263} In this way, code smells can be correctly detected to increase software sustainability. However, only a limited number of architectures were considered and the design aspects like reproducibility, change management, etc which also affect sustainability were given minimal importance.

\section{Sustainability Aspects of User Experience}
\label{ux}

This section highlights the User Experience (UX) perspectives which impact software sustainability from technical standpoint. UX is an important software aspect which can affect sustainability of the software \cite{da2000user}. It can be divided into two sub-categories which are \textit{User Feedback (non-technical)} and \textit{Graphical User Interface (GUI) (technical)} \cite{kitzes2017practice}. We discuss the related literature of the aforementioned sub-categories with respect to software sustainability after defining these terms as follows.

\begin{enumerate}
    \item \textit{User Feedback:} This is the process of analyzing the viewpoints and experience shared by users which provides software engineers an in-depth analysis of the features which are liked by various users, bugs which users want to be fixed \cite{gilb1988principles}. Also it is the collection of information of what the users think is important for longevity of the software \cite{gilb1988principles}. It can include conducting formal surveys on the software, collecting user reviews and informally talking to users via conferences and meetings \cite{gilb1988principles}. Analyzing the user feedback will enable making better quality software which are more acceptable to users and hence contribute to sustainament of the software \cite{gilb1988principles}.
    
    \item \textit{Graphical User Interface (GUI):} It is the interface via which the users interact with the software and its processes \cite{gilb1988principles}. The GUI needs to be user-friendly, and it should adhere to the demands of the users \cite{gilb1988principles}. A software can be very effective, however, if its GUI is not user-friendly then in the long term it will not be used \cite{gilb1988principles}. This is applicable for both closed source and open source software, as shown in the table in this section.
\end{enumerate}

\begin{table*}
\centering
\begin{tabular}{ |p{2cm}|p{5.4cm}|p{5.4cm}|  }
\hline
\multicolumn{3}{|c|}{User Experience} \\
\hline
Sustainability Criteria &Open Source &Closed Source \\
\hline
User Feedback & Standardized approach to integrating stakeholder feedback regarding user friendliness and sustainability requires extensive research \cite{kitzes2017practice}. Also, many contributors may provide many feedbacks which can become uncontrollable to manage and integrate \cite{kitzes2017practice}.  &Qualitative user feedbacks regarding sustainability be integrated by dedicated team of UX engineers \cite{da2000user}. It may not cause issues in monopoly market, however, sustainability of the software may be threatened in a competitive market \cite{7166099}.\\
\hline
Graphical User Interface & Sustainable GUI implementation depends on parametric and associative application, capable of developing easy-to-use and customized GUI \cite{shneiderman2010designing, tidwell2010designing}. & Reliable and reproducible GUI is found in closed software mainly due to the same look and feel used in those \cite{shneiderman2010designing}. This is prevalent in most of the long serving commercial software \cite{shneiderman2010designing}. \\
\hline
\end{tabular}
\caption{User experience requirements for sustainability of software}
    \label{tableUX}
\end{table*}
The user experience and usability attributes of software sustainability have been highlighted in Table \ref{tableUX}. It is seen that taking feedback about what users think of the \textit{GUI} of the software and integrating the useful feedbacks after prioritization is an important criteria to make the software acceptable to the users \cite{kitzes2017practice}. Understanding which \textit{GUI} fits the best requirements and serves the users is key to ensure that the software is used by them over time \cite{7166099}.  As a result, a software needs to have desirable \textit{GUI} if it is to sustain a long period of time.

\subsection{User Feedback}

Best practices in software usability and user experience can have significant effect on software sustainability \cite{kitzes2017practice}. Software failures can be fixed at lower cost, enhance performance and increase productivity of software systems \cite{gilb1988principles}. Importance on \textit{UX} have been emphasized in the academia by designing course works on \textit{Human Computer Interaction (HCI)}. However, most of the \textit{HCI} courses are non major and considered to be esoteric by researchers coming from scientific backgrounds. The dynamic nature of scientific software has made user experience an important criteria for its sustainability \cite{kitzes2017practice}. As a result, Kitzes et al. \cite{kitzes2017practice} combined heuristic studies, participant driven interviews and surveys, usability observations and evaluations to improve the \textit{GUI} of scientific applications. Those experiences have been used to develop sustainable data exploration and analysis methods for \textit{UI}.

When different types of customized \textit{GUI} are available for the same scientific software, it is important to segregate the code of the scientific calculations from the code of \textit{GUI} itself for sustainability \cite{kitzes2017practice}. It has been mentioned earlier that scientific software may be used with different UIs like desktop, mobile, etc and \textit{GUI} of a given platform might need to be customized based on feedback from the users \cite{kitzes2017practice}. Keeping the scientific code engulfed into the code of \textit{GUI} will make this modification and customization difficult to achieve \cite{kitzes2017practice}. Hence, for the purpose of flexibility, the code of the scientific calculations needs to be separately maintained from that of the UIs \cite{kitzes2017practice}. This can be achieved by following a Model View Controller (MVC) architectural pattern. Doing so will ensure that the \textit{GUI} code can be modified and updated for the software over a longer time span, thus contributing towards long time sustainament of software \cite{kitzes2017practice}.

\subsection{Graphical User Interface}

Understanding the behavior of users as there are many software engineering guidelines which promote user-centered design is highly important for software sustainability \cite{shneiderman2010designing}, since it allows the software engineers to identify user requirements regarding \textit{GUI} more specifically. While developing scientific applications which are mostly customized software, it is invaluable to first identify the behavior and working pattern of the users and design the software accordingly, making sure that the users do not need to change their everyday work routines to ensure that the software makes their efforts easier rather than cumbersome \cite{tidwell2010designing}. A friendly \textit{GUI} can ensure that this requirement is met. While designing software, it is important to address the needs of the specific \textit{GUI} users rather than general cases of user requirements \cite{tidwell2010designing}. However, it needs to be addressed that a lot of customized features in the software may also make the users uncomfortable \cite{tidwell2010designing}. Hence, the normal \textit{GUI} features of the software needs to be as is and the exceptional requirements should be integrated \cite{tidwell2010designing}. For scientific applications, it is necessary to address the exceptional requirements more than general purpose software \cite{tidwell2010designing}.

In addition to the above, the user's actions need to be contextualized and their efforts which lead to other tasks need to be perceived correctly \cite{carroll2000making}. This perception should be followed by placing the tasks together and then allowing the tasks to be executed one after the other \cite{carroll2000making}. For this purpose, a design pattern called \textit{Disabled Irrelevant Things} have come into existence which should be considered in this regard \cite{tidwell2010designing}. It causes the \textit{GUI} to display or hide features and options based on their relevance to the user's current or immediate-past action \cite{tidwell2010designing}. In some cases, this prediction might be done by analyzing past experiences of the users on using the software by using artificial intelligence algorithms \cite{tidwell2010designing}. It is found that such design pattern can greatly contribute to software sustainability since they can modify the \textit{GUI} on-the-fly based on user profiles \cite{tidwell2010designing}.

To contradict the above, Buschmann et al.~\cite{buschmann2007pattern} emphasized software architects to think beyond the \textit{GUI} to focus on the fact that despite the role of \textit{GUI} in making software popular to general people, scientific software may require more accuracy at the back-end rather than developing nice \textit{GUI}s. The authors have suggested that scientific software which conduct input and processing of large volumes of data for intelligence will be more productive with a user friendly \textit{CLI} rather than a state-of-the-art \textit{GUI} \cite{buschmann2007pattern}. The \textit{CLI} are better to do quick repetition of complex commands, specially when a command needs to be repeated a number of times with varying parameters as well as other scripting activities \cite{buschmann2007pattern}. 

Additionally, it is seen that many scientific software which are distributed in nature sometimes do not have GUI \cite{buschmann2007pattern}. It is seen that software such as \textit{VisIT} and \textit{Paraview} have renders which will be impractical with a \textit{GUI}. Hence a standard \textit{CLI} is desired at times \cite{buschmann2007pattern}. As a result, many large scientific applications have minimal or no \textit{GUI}. If there is a \textit{GUI}, then the workflow of the \textit{GUI} is separate from that of the back-end computations for the purpose of clarity and ease of software maintenance \cite{buschmann2007pattern}. As a result, it is seen that for scientific software, \textit{CLI}s also play an important role to make them sustainable \cite{buschmann2007pattern}. However, CLI's are another form of interface and the same issue of mixed scientific model and interface can occur, unless MVC is used to provide clear separation.

\section{Non-technical Attributes}
\label{nt}
In this section, we highlight the non-functional aspects of software sustainability. Afterward, we aim to identify the related literature for the sub-categories of non-technical aspects of software sustainability. More specifically, we can divide the non-technical aspects into seven sub-categories which are illustrated below:

\begin{enumerate}
    \item \textit{Effective Documentation:} Effective documentation is the process of writing software design documents and user manual such that they are useful to engineers for understanding, coding and maintaining the software \cite{hutton2016most}. At the same time the user manual should be written in a way so that it is useful to the users to understand the functionality and help them to use the software in the most effective way \cite{katerbow2018recommendations}. Effective documentation can go a long way to ensure that the software is usable for long time, thereby directly contributing to software sustainability.
    
    
    \item \textit{Software Sustainability Manifestos:} This includes the steps, practices, efforts, and necessary interventions which are required to ensure sustainability of a software \cite{Becker:2015:SDS:2819009.2819082}. Here, we try to identify the software sustainability manifestos which have been published in software sustainability conferences and workshops \cite{Becker:2015:SDS:2819009.2819082, Ojameruaye:2016:SDP:2889160.2889218}. These manifestos are useful for \textit{SE}'s to identify the various technical and non-technical software sustainability attributes \cite{Becker:2015:SDS:2819009.2819082}. 
    
    
    \item \textit{Training on Software Sustainability:} It includes capacity building of software engineers regarding the processes or practices which need to be followed to make software sustainable over a longer time period \cite{druskat2018mapping}. It can include training both on the functional and non-functional aspects of software engineering \cite{brown2002reusing}. In this review we identify the literature on the current practices of training both commercial and research software engineers regarding sustainability.
    
    \item \textit{Funding:} It includes the resources for ensuring that the software is sustainable \cite{druskat2016lightning}. Funding is primarily provided to software teams and research institutions for software sustainability practices \cite{8257959}. Later, we identify the papers which state the importance of funding both in the industry and research arena focusing on software sustainability studies.
    
    \item \textit{Leadership Skills:} These are the skills which are required by software project leaders to make sure they plan and design the software in such a way so that those sustain over significant time \cite{Stewart:2015:SSC:2753524.2753533}. To develop such leadership skills, it is important to train the project leaders properly \cite{katerbow2018recommendations}. In this section, we identify the papers which illustrate the relationship between good leadership and software sustainability.
\end{enumerate}

\begin{table*}
\begin{tabular}{ |p{2cm}|p{6.4cm}|p{6.4cm}|  }
\hline
\multicolumn{3}{|c|}{Non-technical factors for sustainability} \\
\hline
Sustainability Criteria & Open Source & Closed Source \\
\hline
Effective Documentation & Standard templates for documentation are seldom followed which affects sustainability. Common tools for documenting open source software are \textit{MarkdownPad, iA Writer, SimpleMDE,} etc \cite{hutton2016most}. &Companies like McAfee, Microsoft, Adobe, etc have their own documentation templates \cite{durdik2012sustainability}. As a result, more structured documentation leads to more sustainable software \cite{durdik2012sustainability}. \\
\hline
Sustainability Manifestos & Open source software do not implement the manifestos till now \cite{jimenez2017four}. However the open source research software provide a suitable platform to implement and test the recently proposed manifestos \cite{morris2009some}.   & Closed source software focus on well established manifestos which are generally older versions of recent proposals for the purpose of reliability \cite{Becker:2015:SDS:2819009.2819082}. Hence they do not contain many new requirements \cite{Becker:2015:SDS:2819009.2819082}. \\
\hline
Training of Different Stakeholders   &Although current training on sustainability focuses on using test cases from successful commercial software, it needs to place enough care in training students in complex software \cite{cabunoc_2018}. The mechanism discussed by the Carpentries \cite{carpentries} can be used here. & Training software engineers to develop sustainable closed source software certainly has its benefits, however they come at an increased cost of finding trainers who know the software well \cite{penzenstadler2011teach}.  \\
\hline
Funding & Primarily suffers from lack of funding \cite{cabunoc_2018}. Although the software is open for all to contribute, many open source software do not get required contributions from society of software engineers \cite{Stewart:2015:SSC:2753524.2753533}. & Generally funding is not a major challenge during initial phase \cite{Stewart:2015:SSC:2753524.2753533}. However, after completion of development the software should generate its own funds by sales revenues to sustain \cite{Stewart:2015:SSC:2753524.2753533}. \\
\hline
Leadership of Project Manager (\textit{PM}) & The sustainability is heavily dependent on leadership of \textit{PM}. The leader needs to envision and address scheduling capability, making the project popular to contributors which is key to sustainability \cite{hettrick2016research, Stewart:2015:SSC:2753524.2753533}. & The leadership skills through which a \textit{PM} handles change management, timeline management, cost management and critical thinking defines the sustainability of the project \cite{plotnick2008leadership}.   \\
\hline
\end{tabular}
\caption{Non-technical requirements for sustainability}
\label{tablePolicy}
\end{table*}

The results of the effects of various policies on sustainability of general software have been highlighted in Table \ref{tablePolicy}. The table provides a comparison of the non-technical activities involved to make open and closed sourced software sustainable. The table identifies that effective, useful documentation is a high priority attribute to preserve software sustainability as too little or too many documentation is not a desirable attribute. Many of the open source software lack any documentation, however their codes may contain comments which can be availed as documentation. Placing correct comments in right places to explain various regions of the code will make changes easier to incorporate and it will enable the software to be updated \cite{huang2018guiding}, which is essential for the long term sustainability of the software. Given proper documentation is available which is the case for most of commercial software, commenting may not be required since there is a frequent need to update the comments whenever the codes are updated \cite{huang2018guiding}. 

\subsection{Effective Documentation}

Documentation and archiving do not guarantee regeneration of experimental results in code~\cite{katerbow2018recommendations}. Hutton et al.~\cite{hutton2016most} suggest researchers to work more closely with research software engineers and focus on the use of open source interfaces. Although their work is based on software for the research area of hydrology, the findings can be generalized. 

A barrier to ensuring sustainability of open source software systems is the lack of good management practices \cite{5602062}. A number of technical and social components in this barrier were identified, one of them is proper documentation \cite{5602062}. The case of patch creations was taken into account and it was stated that lack of proper documentations can lead to severe mismanagement of patches in open source software. They stated that master developers are experts in developing patches, however, they face troubles in maintaining proper documentations \cite{5602062}. The improved documentation of patch creation, release and performance analysis will result in better management of the software, thus increasing the sustainability of critical open source applications.

Effective documentation of software is a critical practice to achieve credible software architecture, reactively eliminate  evolutionary issues, plan variability strategies, manage information, automate software and share knowledge to bring all members of the development team at the same level of understanding the requirements \cite{durdik2012sustainability}. Effective documentation has been identified as one of the key requirements for software maintenance \cite{durdik2012sustainability}. Changes which are accompanied by standardized documentation can be effectively traced even after years of integration into the system \cite{durdik2012sustainability}. Also, the paper stated that documentation can be divided into three categories as \textit{high, medium, and low}.

Lami et al. \cite{6472566} highlighted that effective documentation is a key ingredient to tailoring the processes of software project to meet the peculiarities of the project itself. This is important to ensure the long term sustainability of software projects and prevent requirement decay \cite{6472566}. A high quality document which provides correct information about the required resources is a necessary condition for a sustainable project \cite{6472566}.

\subsection{Sustainability Manifesto}

Renzel et al. presented the concept of \textit{Research Software Engineering} as a new dimension which identifies the requirements for building software for researchers, ensuring sustainability and reliability. The paper identifies the 4 core activities needed to develop sustainable and reliable research software within the context of supporting a successful research software community \cite{jimenez2017four}. The activities are namely \textit{Software Engineering, Community, Training and Policy}. Specifically, \textit{software engineering} is concerned with the process of sustainable software development and includes the people who build it \cite{jimenez2017four}. \textit{Community} provides the platform which can be used by software communities to meet, share and discuss innovative ideas for building sustainable software \cite{jimenez2017four}. 

Morris et al. \cite{morris2009some} identified that \textit{Research Software Engineering} is built to meet a specific research requirement of a research organization, group or individual scientist without much thought on future update and maintenance. \textit{Training} is important to make sure that the software development and maintenance processes remain updated \cite{morris2009some}. The next activity is \textit{Policy} which is about identifying the policy changes required institutionally and culturally and running campaigns to effect that change \cite{morris2009some}. The authors mentioned that in addition to the above, it is also important to make sure that the software becomes a critical aspect and entity of the research team.

The \textit{Karlskrona Manifesto for Sustainability Design} is provided as a platform for communication regarding software sustainability, not limited to the software fraternity \cite{Becker:2015:SDS:2819009.2819082}. It provides a vehicle where the researchers can discuss the myriad of issues regarding sustainability and maintainability, which ultimately contributes to ensuring software quality \cite{Becker:2015:SDS:2819009.2819082}. Lack of information and incorrect information have been identified as key reason for incapability of a software to sustain over time \cite{Ojameruaye:2016:SDP:2889160.2889218}. The authors have provided a decision taking mechanism by modifying the \textit{Cost Benefits Analysis Method (CBAM)} to evaluate architectural design and check the extent to which it is sustainable. In the presence of uncertainty,  this method identifies the sustainability debt \cite{Becker:2015:SDS:2819009.2819082}. Although the system has been tested on a emergency deployment software, the effectiveness of the author's proposal in order to evaluate the architectural sustainability when a large number of third-party components are integrated was not considered.

\subsection{Training Different Actors}

Importance of providing training on sustainability practices in design and implementation of research software guarantees its sustainability \cite{cabunoc_2018}. The researchers have cited that Mozilla Firefox is a successful and sustainable application, mainly because od vast array of training which was provided to the developers. They stated that Mozilla still continues to support, train and research sustainable open source software development practices \cite{brown2002reusing}. The authors highlighted that based on their years of experience in mentoring open source software projects, there are primarily three areas to consider in order to make an open source project sustainable and ensure its growth. This is also supported by the work of the Carpentries \cite{carpentries}, which advocate training on best practices and peer working in the context of a 'community of practice'. The identified areas are:

\begin{enumerate}
    \item \textit{Training on best practices \cite{brown2002reusing}}: Workshops, training and curriculum should be provided and regularly updated to support open source development and software sustainability.
    \item \textit{Peer support \cite{brown2002reusing}}: Working openly is sometimes contradictory to traditional academic practices. Hence there needs to be effective training to build the mindset of peer programming and combined development.
    \item \textit{Resources \cite{brown2002reusing}}: Although open, projects require significant amount of time, manpower and funding to sustain over time. For example, even Mozilla is supported by a number of grants and awards for research software \cite{mozilla.grants}. 
\end{enumerate}

The various institutions worldwide who are concerned with software sustainability have been identified by Druskat et al.  \cite{druskat2018mapping}. The authors stated that these organizations are mostly concerned with scientific software \cite{druskat2018mapping}. They focus on sustainability benchmarks for software designed primarily for researchers and share the issues to the research community via conferences, workshops, etc \cite{druskat2018mapping}. Their findings are used by educational institutions which teach basic programming to its pupils \cite{druskat2018mapping}. Katz et al. \cite{katz2018state} initially provided the diagrammatic representation of the research space. The authors identified the actors at each process of software sustainability and defined their roles. They put the all the actors, their relationships with each other and roles using the \textit{rosusuma} visualization in Python \cite{katz2018state}. Their generated graphs fine grained various broader categories in software sustainability and showed the actors in each category \cite{katz2018state}. As stated in \cite{8257959}, there is a need to train users on how to use software, specially if it is a complex research software requiring specialized efforts. The necessity of training the stakeholders in the software development process was highlighted. 

\subsection{Funding}

Druskat \cite{druskat2016lightning} assumed that most of the software are developed keeping sustainability as a afterthought, mainly due to pressure of short time-to-market and developers are often not educated enough on sustainability improvement techniques. It addresses the fact that economically sustainable software systems need to evolve cost effectively when there is a change in their environment, usage profile and business demands \cite{druskat2016lightning}. The authors mentioned that although many research papers have separately identified the importance of sustainability, there is a need for a holistic catalog of software sustainability guidelines \cite{druskat2016lightning}. 
%
 

\subsection{Leadership}

Professional leadership is needed since it is important to have a transparent control, governance and ownership of the project \cite{plotnick2008leadership}. Open source software does not mean collaboration in all cases of software development \cite{plotnick2008leadership}. However, all projects should be clear on how they can be modified and developed, how the governance policy would work needs to be communicated with the contributors via a dependable communication mechanism. A license should be adopted and any third-party license requirements should be complied \cite{plotnick2008leadership}. A manual needs to be provided to enable other users to modify and use the source code as needed. All those factors need to be managed by a  qualified leader \cite{plotnick2008leadership}. In this way leadership plays an important role towards software sustainability.

Stewart et al.~\cite{Stewart:2015:SSC:2753524.2753533} identified what led to some software applications sustain over the long run. The scope contained software applications which were funded by the US National Science Foundation (NSF) or which were used by researchers funded by the NSF \cite{Stewart:2015:SSC:2753524.2753533}. The paper highlighted that although proper documentation, best engineering practices, proper testing, use of open source technology and licensing were reasons for software to sustain, many applications who followed all these processes resulted in failure. On top of the aforementioned features, a key point which actually resulted in a software being successful involved commanding leadership, dedication of the project owner and regular conferences, meeting, workshops or talks on the project arranged annually \cite{Stewart:2015:SSC:2753524.2753533}. 

Ensuring software sustainability requires a shift in both the research community, software developers, funding agencies and users \cite{hettrick2016research}. \textit{The European Community} \cite{{katerbow2018recommendations}} believes that although a feasible and technical approach is needed to achieve software sustainability, the problem cannot be solved by it alone. The report cites a number of important reasons like software decay, quick shift in technology, societal and cultural barriers as the primary reasons in the lack of sustainability in many research software of today \cite{katerbow2018recommendations}. Career path of software experts, importance of software in research, understanding licensing procedures of research software and clear incentives and impact are the some of the important hindrances in research software engineering \cite{katerbow2018recommendations}. However, how funding can combine with technical attributes of sustainability to provide system wide sustenance is an area of future research interest.

\section{Conclusion}
\label{conclude}

This literature review highlighted that ensuring sustainability of software is an important area of research in software engineering with applications in many diverse fields. The most important application of software sustainability in today's world is that research and commercial organizations like corporates, financial institutions, government offices, etc are moving towards complete automation and they need their software to last a long time. The advantages of sustainability are increased reliability, following industry standard SDLC, opportunity of software evolution, increased efficiency and reduced cost for software clients in the long run as identified here.

In the early and mid 2000s, authors focused on making software secure and reliable. From 2016 and later, the focus changed towards ensuring sustainability of software \cite{penzenstadler2014safety}. This is mainly due to the fact that software nowadays are used for newer purposes and dimensions, many of which require the software to survive for a long time under increased workload. In addition, importance of making cloud based services more sustainable was highlighted in \cite{chang2010review}. Today, all critical functions in an organization are reliant on automation, thus requiring sustainable software. Additionally, the importance of sustainability is so high in the research community that software sustainability institutes have been set up in USA \cite{urssi} and Europe \cite{essi} respectively, which focuses primarily on sustainability of research software.

Considering the factors mentioned above, here we identified the contributions of existing research endeavours covering various technical and non-technical aspects of software sustainability. We discussed the objectives and methodologies of those and identified the limitations. We concluded that technical and non-technical factors cannot provide sustainability of software single-handedly. Hence, a combined implementation of both technical and non- technical factors were needed to achieve sustenance. Also, comparison of open source and closed source software on various factors of sustainability showed that open source software ran the risk of decay caused by lack of continuous funding which threatened sustainability. This provided the opportunity to test new and more effective methods of ensuring sustainability and overcome the aforementioned challenge. Closed source software were found to follow well-established, albeit older methods of sustainability and as a result lacked many new practices.

As highlighted in the previous sections of this paper, the limitations of existing research on sustainability are mainly the lack of sustainability benchmarks and practices at the software design level. Also, application of the methodologies on real-life software requires implementation and performance evaluation. Additionally, how to achieve sustainability of largely distributed software used both by researchers and commercial organizations is an area of significant research interest mainly because those software may be dependent on many un-sustainable third-party APIs. At the same time, providing performance measuring frameworks to calculate the sustainability of a software is an area which requires exploration to a greater detail. Finally, we have seen that many complex research applications lack sustainable design, hence a framework to evaluate, explore and implement sustainability features in those is an area of future research interest.
\section*{References}

\bibliography{mybibfile}

\end{document}